\pdfoutput=1
\documentclass[a4paper,fleqn,usenatbib]{mnras}

\usepackage{newtxtext,newtxmath}

\usepackage[T1]{fontenc}
\usepackage{ae,aecompl}

\usepackage{graphicx}	\usepackage{amsmath}	\usepackage{amssymb}	\usepackage{pdfpages}
\usepackage{gensymb}
\usepackage{tabularx}
\usepackage{subfig}
\usepackage[figuresleft]{rotating}
\usepackage{wrapfig}
\usepackage[outdir=./]{epstopdf}
\usepackage[nice]{nicefrac}
\usepackage{pdflscape}
\newcommand{\angstrom}{\textup{\AA}}

\title[Metallicity and Ionization Mapping]{The SAMI Galaxy Survey: Spatially Resolved Metallicity and Ionization Mapping}

\author[Poetrodjojo et al.]{Henry Poetrodjojo,$^{1,2}$\thanks{E-mail: henry.poetrodjojo@anu.edu.au }
Brent Groves,$^{1,2}$
Lisa J. Kewley,$^{1,2}$
Anne M. Medling,$^{1,3,4}$
\newauthor
Sarah M. Sweet,$^{2,5}$
Jesse van de Sande,$^{6}$
Sebastian F. Sanchez,$^{7}$
Joss Bland-Hawthorn,$^{6}$
\newauthor
Sarah Brough,$^{8}$
Julia J. Bryant,$^{6,9,10}$
Luca Cortese,$^{2,11}$
Scott M. Croom,$^{6,10}$
\newauthor
\'Angel R. L\'opez-S\'anchez,$^{2,9,12}$
Samuel N. Richards,$^{13}$
Tayyaba Zafar,$^{9}$
Jon S. Lawrence,$^{9}$
\newauthor
Nuria P.F. Lorente,$^{9}$
Matt S. Owers$^{9,12}$
and Nicholas Scott$^{6}$
\\
$^1$Research School of Astronomy and Astrophysics, The Australian National University, Canberra, ACT 2611, Australia\\
$^2$ARC Centre of Excellence for All Sky Astrophysics in 3 Dimensions (ASTRO 3D)\\
$^{3}$Cahill Center for Astronomy and Astrophysics California Institute of Technology, MS 249-17 Pasadena, CA 91125, USA\\
$^{4}$Hubble Fellow\\
$^5$Centre for Astrophysics and Supercomputing, Swinburne University of Technology, PO Box 218, Hawthorn, VIC 3122, Australia\\
$^6$Sydney Institute for Astronomy (SIfA), School of Physics, University of Sydney, NSW 2006, Australia\\
$^{7}$Instituto de Astronom\'ia, Universidad Nacional Aut\'onoma de M\'exico, A.P. 70-264, 04510, M\'exico, D.F\\
$^8$School of Physics, University of New South Wales, NSW 2052, Australia\\
$^9$Australian Astronomical Observatory, 105 Delhi Rd, North Ryde, NSW 2113, Australia\\
$^{10}$ARC Centre of Excellence for All-sky Astrophysics (CAASTRO)\\
$^{11}$International Centre for Radio Astronomy Research, University of Western Australia, 35 Stirling Highway, Crawley, WA 6009, Australia\\
$^{12}$Department of Physics and Astronomy, Macquarie University, NSW 2109, Australia\\
$^{13}$SOFIA Operations Center, USRA, NASA Armstrong Flight Research Center, 2825 East Avenue P, Palmdale, CA 93550, USA
}
\date{Accepted XXX. Received YYY; in original form ZZZ}

\pubyear{2018}

\begin{document}
\newcommand{\rtt}{$\textrm{R}_{23}$ }

\newcommand{\rtteq}{$(\textrm{[OII]} \lambda3726, \lambda3729+\textrm{[OIII]} \lambda4959, \lambda5007)$/$\textrm{H}\beta$ }

\newcommand{\mt}{$M_{20}$ }

\newcommand{\htwo}{H{\sc ii} }

\newcommand{\bpt}{Baldwin, Phillips and Terlevich (BPT) diagram \citep{baldwin81}}

\newcommand{\stha}{[SII]/H$\alpha$ }

\newcommand{\ntha}{[NII]/H$\alpha$ } \label{firstpage}
\pagerange{\pageref{firstpage}--\pageref{lastpage}}
\maketitle

\begin{abstract}
We present gas-phase metallicity and ionization parameter maps of 25 star-forming face-on spiral galaxies from the SAMI Galaxy Survey Data Release 1. Self-consistent metallicity and ionization parameter maps are calculated simultaneously through an iterative process to account for the interdependence of the strong emission line diagnostics involving ([OII]+[OIII])/H$\beta$ (\rtt) and [OIII]/[OII] (O32). The maps are created on a spaxel-by-spaxel basis because \htwo regions are not resolved at the SAMI spatial resolution. We combine the SAMI data with stellar mass, star formation rate (SFR), effective radius (R$_e$), ellipticity, and position angles (PA) from the GAMA survey to analyze their relation to the metallicity and ionization parameter. We find a weak trend of steepening metallicity gradient with galaxy stellar mass, with values ranging from -0.03 to -0.20 dex/R$_e$. Only two galaxies show radial gradients in ionization parameter. We find that the ionization parameter has no significant correlation with either SFR, sSFR (specific star formation rate), or metallicity. For several individual galaxies we find structure in the ionization parameter maps suggestive of spiral arm features. We find a typical ionization parameter range of $7.0 < \log(q) < 7.8$ for our galaxy sample with no significant overall structure. An ionization parameter range of this magnitude is large enough to caution the use of metallicity diagnostics which have not considered the effects of a varying ionization parameter distribution.
\end{abstract}


\section{Introduction}\label{sec:Introduction}
The accurate measurement of gas-phase metallicity and ionization parameter in galaxies is becoming increasingly essential as we probe deeper into the universe and observe galaxies at high redshift. The gas-phase metallicity is strongly affected by processes that occur during the evolution of galaxies such as gas inflows, galaxy mergers, and galactic winds. Because of this connection, the distribution of the metallicity in galaxies provides a strong constraint on their growth and formation and recent dynamical processes.

\par

Simulations by \citet{pilkington12} show that a negative gas-phase metallicity gradient provides strong evidence for inside-out disc formation \citep{matteucci89,boissier99}. In this model, a negative metallicity gradient implies that the central metal-rich gas has been forming stars for longer than the metal poor outskirts.
\par

Local disk galaxies typically have a negative metallicity gradient \citep{zaritsky94,moustakas10,rupke10,sanchez14}. A dependence on morphology was observed by \citet{vila-costas92,zaritsky94,martin94}, in that barred galaxies have shallower metallicity gradients than unbarred galaxies.

\par

Large-scale gas inflows can disrupt metallicity gradients. \citet{kewley10} showed that the metallicity gradients of close pair galaxies are significantly shallower than those of isolated galaxies. Tidal disruptions from  galaxy interactions drive pristine gas from the outskirts into the central regions, diluting the metal-rich centre. \citet{lopez-sanchez15} showed that one of the spiral arms belonging to NGC 1512 had a flattened metallicity gradient due to its interaction with nearby dwarf galaxy NGC 1510. \citet{sanchez14} also found significantly flatter metallicity gradients in galaxies that show signs of merger activity.

\par

With advances in integral field spectroscopy (IFS), we can now spatially map the metallicity across galaxies, allowing for a deeper insight into azimuthal and radial variations within a galaxy. Several small scale surveys such as PPAK IFS Nearby Galaxies Survey (PINGS) \citep{rosales-ortega10}, the VIRUS-P Investigation of the Extreme Environments of Starbursts (VIXENS) \citep{heiderman11} and the VIRUS-P Exploration of Nearby Galaxies (VENGA) \citep{blanc13,kaplan16} have been conducted. The first large survey was the Spectrographic Areal Unit for Research on Optical Nebulae (SAURON) survey \citep{dezeeuw02}, which initially observed 72 low redshift early-type galaxies (ETG) using IFS technology, and was later continued into the $\textrm{ATLAS}^{\textrm{3D}}$ survey \citep{cappellari11}, observing 260 galaxies at $z<0.01$.

\par
The Calar Alto Legacy Integral Field Area Survey (CALIFA) survey \citep{sanchez12} consists of 600 galaxies with $z<0.03$. \citet{sanchez14} used $\sim$306 CALIFA galaxies to analyse the oxygen abundance gradients in galaxy disks and found that all undisturbed galaxies with a disk presented similar radial metallicity gradients when normalised to the size of the disk. They showed that the existence of a characteristic metallicity gradient is independent of luminosity, mass and morphology when normalised to the size of the disk.
\par
Similar results were obtained by \citet{sanchez12} using PINGS data and \citet{ho15,sanchez-menguiano16} who both used CALIFA data for their analysis. This contradicts the findings of \citet{vila-costas92,zaritsky94,martin94} who found a clear variation in metallicity gradient between barred and unbarred galaxies. These differences could be due to earlier studies using a smaller sample size \citep{ho15} or inconsistent methods of measuring metallicity gradients \citep{sanchez-menguiano16}. 

\par

 While the $\textrm{ATLAS}^{\textrm{3D}}$ and the CALIFA surveys have now managed to amass hundreds of galaxies, they do not have the multiplexing technology to easily reach thousands of galaxies. This was made possible by the development of the hexabundle \citep{bland-hawthorn11} which led to the development of the Sydney-AAO Multi-object Integral field (SAMI) spectrograph \citep{croom15}. The SAMI Galaxy Survey \citep{bryant15} will complete in 2018 with 3600 galaxies across a wide range of environments and stellar masses, allowing for the disentanglement of degeneracies. This will be followed by the Hector survey with an order of magnitude increase in the observed number of galaxies \citep{bland-hawthorn15}. 
 
 \par
 The Mapping Nearby Galaxies at Apache Point Observatory (MaNGA) survey \citep{bundy15} is an ongoing galaxy survey aiming to achieve spatially resolved spectra of 10,000 nearby galaxies. MaNGA uses specially designed fibre bundles \citep{drory15} that vary in diameter and number of fibres to allow the observation of a representative sample of local galaxies in the mass range $10^{9}<M/M_{\odot}<10^{12}$. Fibre bundles range from 19-127 fibres with an on-sky diameter ranging from $12\arcsec - 32\arcsec$.
 \par
 With a sample of 550 galaxies from the MaNGA survey, \citet{belfiore17} found a steepening of the metallicity gradients with stellar mass up to a mass of $\log(M_{*}/M_{\odot}) < 10.5$. For more massive galaxies, the metallicity gradient flattens slightly as the metallicity of the galaxy reaches a constant value.

\par
The gas-phase metallicity is most commonly presented as the ratio between the abundance of oxygen, the most abundant heavy element by mass, and hydrogen. For star-forming galaxies, the metallicity is usually determined using the ratios of the strong emission lines. Some of the popular strong emission line diagnostics include \rtteq \citep[\rtt;][hereafter KK04]{kobulnicky04}, $\textrm{[NII]}\lambda6583$/$\textrm{[OII]}\lambda3726, \lambda3729$ \citep[N2O2;][hereafter KD02]{kewley02}, $(\textrm{[OIII]}\lambda5007$/$\textrm{H}\beta)$/$(\textrm{[NII]}\lambda6583$/$\textrm{H}\alpha)$ \citep[O3N2;][hereafter PP04]{pettini04}, $\textrm{[NII]}\lambda6583$/$\textrm{H}\alpha$ \citep[N2HA;][]{pettini04} and $\textrm{[NII]}\lambda6583$/$\textrm{[SII]}\lambda6717, \lambda6731$ \citep[N2S2;][hereafter D16]{dopita16}. Each diagnostic has its own set of advantages and disadvantages making them suitable for different situations. These diagnostics are then calibrated against data to determine metallicities. However, all these metallicity calibrations are inconsistent with each other, leading to different abundances depending on the particular diagnostic and calibration used. \citet{kewley08} attempts to consolidate the many metallicity diagnostics and calibrations by providing conversion polynomials between them. For a comprehensive review and analysis of the various metallicity diagnostics and calibrations, see \citet{kewley08,lopez-sanchez12}. 

\par
Ionization parameter strongly affects many metallicity diagnostics (eg. N2HA, O3N2, \rtt). The ionization parameter is defined as:

\begin{equation}
	\begin{centering}
		q=\frac{S_{H^{0}}}{n}
		\label{qequation}
	\end{centering}
\end{equation}
where $S_{H^{0}}$ is the ionizing photon flux per unit area and $n$ is the number density of the interstellar medium. The ionization parameter is a measure of the amount of ionizing photons passing through the interstellar medium per hydrogen atom. \citet{dopita14} found a strong correlation between ionization parameter and star-formation rate (SFR) and suggest that the correlation is caused by the change in geometry of the molecular and ionized gas with environment. Similar results are obtained by \citet{kaplan16}, who found strong evidence of the existence of radial ionization parameter gradients and a correlation with SFR.

\par
The dependence of metallicity diagnostics on ionization parameter is clearly shown in \citet{lopez-sanchez11,ho15}. The KD02 N2O2 diagnostic is relatively independent of ionization parameter but the PP04 O3N2 diagnostic was empirically calibrated without taking into account the effect of ionization parameter. \citet{ho15} showed that the differences between the two diagnostics correlates strongly with the ionization parameter, highlighting the importance of correcting for ionization parameter when calculating metallicity.
 
\par
In this paper we simultaneously constrain the metallicity and ionization parameter of pure star-forming SAMI galaxies through an iterative process and produce self-consistent spatially resolved metallicity and ionization parameter maps. We derive metallicity gradients and analyse the spatial distribution of the ionization parameter. We confirm the results of \citet{sanchez12,sanchez14,ho15,sanchez-menguiano16} by obtaining consistent metallicity gradient values. We find a weak mass-dependence of metallicity gradients using the KK04 \rtt metallicity diagnostic, showing a similar trend to \citet{belfiore17}. We show that the ionization parameter does not change as a function of radius with most star-forming galaxies and we investigate whether the ionization parameter correlates with fundamental galaxy properties like metallicity, SFR and specific star formation rate (sSFR). Finally we show the implications of excluding the ionization parameter from metallicity calculations.
\par
We structure this paper in the following way. Section 2 describes the SAMI Galaxy Survey and how we select our sub-sample from the data available. We outline the methods we use for determining the metallicity and ionization parameter while taking into account the interdependence of the diagnostics in Section 3. In Sections 4 and 5, we present and briefly compare to previous work, our results of the metallicity and ionization parameter analysis respectively. We discuss the results and provide a summary and conclusion in Sections 6 and 7. Throughout the entire paper, we assume the following values for cosmological constants, $H_{0} = 70 \textrm{km s}^{-1} \textrm{Mpc}^{-1}$, $\Omega_{\textrm{M}}=0.3$ and $\Omega_{\Lambda}=0.7$.
 
\section{Sample Selection}
\subsection{SAMI Galaxy Survey}
The SAMI Galaxy Survey \citep{croom12} is an ongoing integral field spectroscopic survey of $\sim3600$ low-redshift (z<0.12) galaxies primarily selected from the Galaxy and Mass Assembly (GAMA) survey \citep{driver11}, with the addition of 8 galaxy clusters to extend the sampling of environmental density \citep{owers17}. The survey uses the SAMI spectrograph on the 3.9 metre Anglo-Australian Telescope at Siding Spring Observatory. The final primary survey targets consist of galaxies with stellar masses between $10^{7} - 10^{12}M_{\odot}$, redshifts between $0.004 < z < 0.095$ and magnitudes $r_{pet} < 19.4$ mag. For full details on the SAMI Galaxy Survey selection, refer to \citet{bryant15}.
\par
The SAMI data are released as a red and blue data cubes for each galaxy, with $50\times50$ 0.25 ($0.5\times0.5$) arcsec$^2$ spaxels covering the 14.7\arcsec diameter aperture of the SAMI hexabundle and an average seeing of 2.16\arcsec (see \citealt{green17} for details). The blue cube covers a wavelength range between $3700-5700\angstrom$ with a spectral resolution of R=1812 and the red cube covers a wavelength range between $6300-7400\angstrom$ with a spectral resolution of R=4263 \citep{vandesande17}. These spectral ranges cover the strong optical emission lines commonly used as diagnostics of the gas-phase metallicity and ionization parameter: [OII]$\lambda3726, \lambda3729$, H$\beta\lambda4861$, [OIII]$\lambda5007$, H$\alpha\lambda6563$, [NII]$\lambda6583$ and [SII]$\lambda6717, \lambda6731$. The red and blue datacubes are analysed using LaZy-IFU (LZIFU v0.3.2); \citep{ho16}. LZIFU extracts total line fluxes for the dominant emission lines by fitting and subtracting the underlying continuum and then fitting the dominant emission lines using up to three Gaussian profiles. LZIFU returns maps of the flux and flux errors for each emission line, as well as maps of the ionized gas velocity and velocity dispersion and their associated errors (see \citet{ho16} for a detailed explanation of the routine).

The galaxy sample for which we determine the resolved metallicity and ionization parameter is based on the 772 galaxies in Data Release 1 of the SAMI Galaxy Survey \citep{green17}. However, to obtain the highest S/N and largest possible maps of these parameters, we placed the following selection criteria on the galaxies (each of which is elaborated in the following subsections):

\begin{itemize}
	\item Star-forming galaxies free of AGN and shocks using the \citet{kewley06} classification scheme
	\item Emission-line maps covering at least 70$\%$ of the hexabundle field of view in all emission lines used
	\item Face-on galaxies with an inclination angle less than 60 degrees based on measurements from the GAMA survey
	\item Each galaxy is sampled to at least 1 effective radius ($R_{\rm e}<7.4\arcsec$) based on measurement from the GAMA survey
	\item A S/N ratio > 3 in the [OII], H$\beta$, [OIII], H$\alpha$, [NII] and [SII] emission line fluxes for each spaxel
\end{itemize}

These selection criteria limit our sample to 25 star-forming, 'best-case' scenario galaxies to determine reliable metallicity and ionization parameter maps. The final sample of galaxies and their global properties as defined in the GAMA galaxy catalogue are given in Table 1. We use the R-band effective radii throughout this study. We also give the H$\alpha$ derived SFR assuming a Salpeter \citep{salpeter55} initial mass function (IMF) as well as stellar mass derived from the mass-luminosity relation \citep{taylor11}. For a comparison between the SFR values determined with GAMA data and SAMI data, see \citet{medling18}. 
\par
In future studies, we intend to expand this analysis to the full SAMI Galaxy Survey sample. With a larger sample, we will probe the relationships between metallicity and ionization parameter with galaxy properties in greater detail. 
\subsection{Star-Forming Galaxies}
For typical blue cloud galaxies, strong emission lines arise predominantly from \htwo regions surrounding recently formed massive stars. However, emission lines can also arise from gas excited by other sources of ionization, such as shocks or Active Galactic Nuclei (AGN) e.g. \citet{groves04}. The metallicity and ionization parameter diagnostics that we rely on are calibrated assuming \htwo region emission, and cannot be simply applied to galaxies with significant contribution from other ionizing sources to the emission lines. In some cases it is possible to separate the star-formation dominated and other ionizing sourced line emission \citep[eg.][]{davies14,davies16}, but in our case we chose to remove all galaxies that showed significant non-star-forming emission.
\par
\citet{medling18} created star formation masks for the SAMI galaxy survey DR1 using the classification scheme of \citet{kewley06}, that uses strong emission line ratios to create diagnostic curves that distinguish when non-star-forming emission is present:

\begin{equation}
\log{\frac{\textrm{[OIII]}}{\textrm{H}\beta}}>\frac{0.61}{\log{\frac{\textrm{[NII]}}{\textrm{H}\alpha}}-0.05}+1.30,
\label{NIIClass}
\end{equation}

\begin{equation}
\log{\frac{\textrm{[OIII]}}{\textrm{H}\beta}}>\frac{0.72}{\log{\frac{\textrm{[SII]}}{\textrm{H}\alpha}}-0.32}+1.30,
\label{SIIClass}
\end{equation}

\begin{equation}
\log{\frac{\textrm{[OIII]}}{\textrm{H}\beta}}>\frac{0.73}{\log{\frac{\textrm{[OI]}}{\textrm{H}\alpha}}+0.59}+1.33.
\label{OIClass}
\end{equation}

Spaxels with a S/N > 5 in the emission line fluxes that satisfy any (fail all of) these criteria were classified as non-star-forming (star-forming). In the case where S/N < 5, \citet{medling18} used a conservative approach to ensure that the sample remained clean.

\par
After identifying the dominant ionization mechanism in each spaxel, \citet{medling18} calculated the fraction of the hexabundle field of view which is filled by the star forming spaxels. For our analysis we require that $70\%$ of the hexabundle was star-forming to ensure that a significant portion of the field of view is filled. This reduces our DR1 SAMI galaxy sample to 91 galaxies.
\par
Implementing this sample selection cut excludes galaxies based on several other galaxy properties. This cut clearly removes galaxies belonging to the red sequence, leaving only galaxies that lie within the blue cloud. However, we are also performing cuts based on angular size and ellipticity. Since we require at least $70\%$ of the hexabundle to be filled with star-forming spaxels, we remove both small blue galaxies as well as highly inclined galaxies which do not sufficiently fill the field of view. 
\par
92 galaxies have star-formation fractions (fraction of H$\alpha$ spaxels classified as star-forming) less than $10\%$. This subset is filled with red sequence galaxies that no longer undergo significant star-formation. Of the remaining 680 blue cloud galaxies, 151 ($22\%$) galaxies have star-formation fractions greater than $70\%$. For 91 ($13\%$) galaxies, the star-forming spaxels also fill $70\%$ of the total hexabundle field-of-view. Although 60 galaxies have star-formation fractions greater than $70\%$, their angular size is either too small or are too inclined to fill the hexabundle field of view. 
\par
Overall this cut removes non-star-forming elliptical galaxies as well as AGN and shock-dominated galaxies, where the majority of spaxels satisfy the diagnostic curves shown in Equations \ref{NIIClass}, \ref{SIIClass} and \ref{OIClass}. Although small low surface brightness galaxies have high star formation fractions with respect to their size, their angular size is not large enough to sufficiently fill the hexabundle, making it difficult to derive radial gradients.
\subsection{Well-resolved radial profiles}
To measure reliable radial metallicity gradients, we require well-sampled radial profiles of the emission-line fluxes. In practice, this means that we select galaxies with inclinations of < $60\degree$ and effective radii $R_{\rm e} < 7.4\arcsec$ for face-on galaxies to ensure that we sample at least 5 resolution elements across $1R_{\rm e}$ and that we limit confusion along the minor axis. These selection criteria further reduce our galaxy sample to 38 galaxies.

\subsection{High S/N Galaxies}
To obtain reliable metallicity and ionization parameter measurements, we require spaxels to have a S/N$>3$ in all of the emission line fluxes used in our diagnostic ratios: [OII]$\lambda3726, \lambda3729$, H$\beta\lambda4861$, [OIII]$\lambda5007$, H$\alpha\lambda6563$, [NII]$\lambda6583$ and [SII]$\lambda6717, \lambda6731$. We applied this criterion to all spaxels in our remaining galaxy sample, while still requiring a coverage of 70\% of the SAMI field-of-view. This final cut, especially the limit on [OII], reduced our sample to 28 galaxies. A further 3 galaxies had such a small redshift such that the [OII] emission line was not redshifted enough to fall in the range of the detector. 
\par
\begin{figure*}
	\centering
	\includegraphics[width=\textwidth]{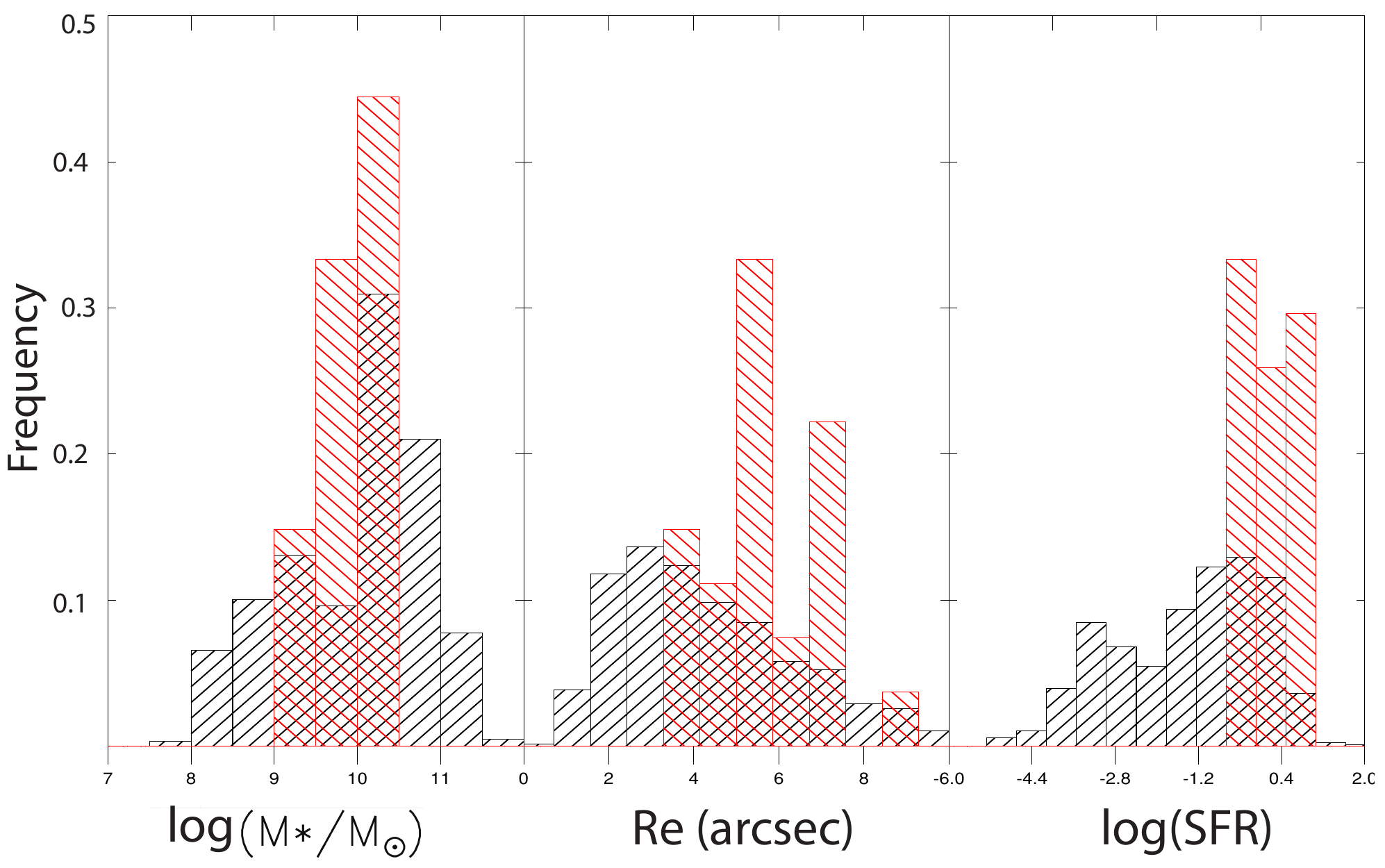}
	\caption{
	Comparison between DR1 of the SAMI Galaxy Survey (Black) and the final galaxy sample used for our analysis (Red). \textit{Left Panel:} We have selected galaxies in the middle of the mass range ($9.0 <\log{M_{*}/M_{\odot}}<10.5$) of DR1. Less massive galaxies are selected to have smaller redshifts and have comparable effective radii. This means that they are more diffuse and harder to obtain reliable S/N, leading to the lower mass limit. The upper mass limit is due to our restriction on sampling to at least 1Re. As we are observing a fairly narrow redshift range, more massive galaxies tend to have a larger apparent size, meaning we are unable to achieve the minimum 1Re we desire. \textit{Middle Panel:} Again we sample the middle range of effective radii for the same reasons as we sample the middle range of stellar mass. The only exception is a slight spike beyond Re $> 8\arcsec$. This comes from GAMA-422366, which has an ellipticity of 0.354, allowing it to be sampled beyond 1Re along the minor axis despite the effective radius being larger than the SAMI field of view radius. \textit{Right Panel:} Since we are aiming to only look at galaxies with high SFR fractions, we are only sampling the high SFR end of the DR1 SAMI Galaxy Survey. 
	}
	\label{sample_histogram}
\end{figure*}

Applying this final cut reduces the final sample to 25 face-on resolved star forming galaxies. Figure \ref{sample_histogram} compares our sample to all the galaxies in DR1 of the SAMI galaxy survey. It is clear that our sample is extremely biased with respect to the SAMI galaxy survey. The low-mass galaxies have an effective radii distribution similar to the whole DR1 sample. Since these low-mass galaxies are spread over the same area as higher-mass galaxies, they are more diffuse and hence harder to detect to a reliable S/N. The S/N requirements outlined in \citet{medling18} mean that low S/N spaxels are usually classified as non-star-forming, causing the lower-mass limit. The upper-limit of about $\log(M_{*}/M_{\odot})=10.5$ is due to the fact that the blue sequence turns over at $\log(M_{*}/M_{\odot})\approx10.5$ \citep{karim11}, with more massive galaxies belonging to the non-star-forming red sequence. We sample the middle range of effective radii due to our requirements on sampling to at least 1R$_{e}$ and filling $70\%$ of the hexabundle. We would not be able to sufficiently sample large angular size galaxies out to 1R$_{e}$ and small angular size galaxies would not cover enough of the hexabundle. We have purposely selected galaxies to have high SFR, leading to the extreme bias towards high SFR galaxies compared to the DR1 sample. Since smaller galaxies tend to be the one with low SFR, by removing all the low mass galaxies, we are left only with very high star-forming galaxies.

\setlength\tabcolsep{1.5pt}
\begin{table}
\centering
\begin{tabular}{l*{10}{c}}
GAMA& RA & Dec & z & $\log$(Mass) & SFR & R$_{e}$ &Elip&PA \\
&	deg&	deg&	&		$M_{*}/\textrm{M}_{\odot}$&	$\nicefrac{\textrm{M}_{\odot}}{\textrm{yr}}$&	arcsec& 1-(b/a) &  \\
\hline
008353&182.0164&0.6976&0.020&9.35&0.51&5.37&0.373&58.9\\
022633&178.4447&1.1934&0.070&10.28&9.93&5.08&0.297&107.3\\
030890&177.2579&-1.1025&0.020&9.79&0.76&7.56&0.435&27.1\\
053977&176.0183&-0.2109&0.048&9.94&5.02&3.79&0.202&103.7\\
077754&214.6477&0.1577&0.053&10.47&9.19&7.03&0.438&81.2\\
078667&218.0908&0.1781&0.055&10.14& - &6.85&0.225&19.7\\
084107&175.9984&0.4280&0.029&9.62&0.60&5.05&0.231&77.4\\
100192&185.9276&0.9621&0.024&9.33&0.18&5.66&0.080&127.3\\
106717&217.0188&1.0063&0.026&10.16&3.25&5.23&0.145&153.6\\
144402&179.9611&-1.3819&0.036&10.25& - &4.14&0.296&23.4\\
184415&176.3419&-1.5652&0.028&9.54&0.50&3.62&0.352&134.7\\
209181&132.1251&0.1708&0.058&10.24&3.71&4.31&0.442&120.6\\
209743&134.6767&0.1914&0.041&10.16&2.15&6.95&0.137&10.1\\
220439&181.6315&1.6166&0.019&9.52&0.72&5.64&0.237&7.1\\
227970&215.6045&1.1976&0.054&10.12&3.47&4.36&0.122&90.0\\
238395&214.2431&1.6404&0.025&9.88&2.18&4.11&0.341&157.9\\
273952&185.9555&1.3751&0.027&9.57&0.08&6.68&0.230&67.2\\
279818&139.4387&1.0554&0.027&9.55&0.58&7.24&0.476&40.0\\
422366&130.5955&2.4973&0.029&9.64&0.41&8.86&0.354&168.7\\
463288&212.4848&-1.2400&0.025&9.63&2.48&7.26&0.183&121.6\\
487027&222.6791&-1.7148&0.026&10.11&9.04&6.22&0.408&31.6\\
492414&216.5031&-1.4117&0.055&10.06&1.39&4.40&0.240&110.9\\
610997&182.8690&0.3786&0.020&9.32&0.21&5.48&0.043&179.6\\
618116&214.4055&0.3290&0.051&10.24&2.16&5.76&0.181&166.5\\
622744&134.8299&0.7977&0.013&9.07&1.42&5.39&0.227&52.4\\	
\end{tabular}
\caption{Selected sample of galaxies from the SAMI galaxy survey and their properties used for our analysis, obtained from the GAMA survey. SFR was not available for GAMA-78667 and GAMA-144402.}
\end{table}
\setlength\tabcolsep{6pt} 
\section{Determining Metallicity and Ionization Parameter}
\subsection{Extinction Correction}
We first correct the emission lines for the attenuation by dust in the interstellar medium (ISM). The attenuation of emission lines is wavelength dependent, meaning that emission line diagnostics that use emission lines with wide wavelength differences are most heavily affected. To extinction correct the emission lines, we create maps of the observed Balmer ratio, (H$\alpha$/H$\beta)_{\rm obs}$. We solve for $E(B-V)$ by using the relation:
\begin{equation}
	E(B-V)=\log(\frac{(\textrm{H}\alpha/\textrm{H}\beta)_{\rm obs}}{(\textrm{H}\alpha/\textrm{H}\beta)_{\rm int}})/{0.4[k(\rm H\beta)-k(\rm H\alpha)]}
	\label{dustext}
\end{equation}
where $(\textrm{H}\alpha/\textrm{H}\beta)_{int}$ is the intrinsic ratio of 2.86 \citep{osterbrock89} assuming case B recombination. We use the \citet{cardelli89} extinction curve and assume a typical $R(V)$ value of 3.1 to determine $k$ values for H$\alpha$ and H$\beta$. We then use the calculated $E(B-V)$ to determine A($\lambda$) at our emission line wavelengths to de-redden the emission line fluxes.
\subsection{Aliasing caused by Differential Atmospheric Refraction}
As described in \citet{green17}, differential atmospheric refraction (DAR) can combine with limited spatial sampling as done in the SAMI survey to create aliasing effects on the spectra. The aliasing is caused by the atmostphere and is made worse by the way the SAMI instrument performs its drizzling to fill in gaps between fibres. While the overall DAR shift is accounted for, a combination of the seeing and sampling in the SAMI survey has meant the DAR has introduced aliasing into the spectra on scales comparable to the point spread function (PSF). This aliasing is most noticeable when taking the ratio of two widely separated wavelength emission lines. With an oversampled PSF, we expect variations between neighboring spaxels to be normally distributed. However, with aliasing, we find excess noise in flux ratios with wide wavelength separations. To correct for what is in effect a variation of the PSF with wavelength, when examining the Balmer decrement, \citet{medling18} smoothed the line ratio map by using a 5x5 spaxel Gaussian kernel with a full-width at half maximum (FWHM) of 1.6 spaxels ($0.8\arcsec$). This smoothing brings the noise down to levels we would expect with an oversampled PSF. We apply this same method not only to our Balmer decrement, but also to our metallicity and ionization parameter diagnostics (\rtt, N2O2, O32), as these all have a significant wavelength gap between emission lines.
\subsection{\rtt Diagnostic}
One of the most popular and well calibrated strong emission line metallicity diagnostics is \rtteq, also known as \rtt, first introduced by \citet{pagel79}. This diagnostic measures the sum of the two dominant ionization states of oxygen in H{\sc ii} regions, which captures the majority of the element. However, this diagnostic is sensitive to temperature and ionization, which has resulted in many \rtt calibrations, each leading to different metallicity estimates \citep{pagel79,pagel80,edmunds84,mccall85,dopita86,torres-peimbert89,mcgaugh91,zaritsky94,pilyugin00,charlot01,kewley02,kobulnicky04}. For a comprehensive review and analysis of various metallicity calibrations see \citet{kewley08}. Furthermore, due to this sensitivity to temperature, the \rtt diagnostic can be degenerate with both a high and low-metallicity solution.
\par
Some \rtt metallicity diagnostics take the ionization parameter into account \citep{mcgaugh91,kewley02,kobulnicky04}. However, the determination of the ionization parameter is similarly difficult because many ionization parameter diagnostics have a significant dependence on metallicity. By using an iterative method described in KD02, we are able to constrain metallicity and ionization parameter simultaneously (see Section 3.5). 
\subsection{O32 Diagnostic}
One way of measuring the ionization parameter is to measure the relative flux of emission lines from high and low-ionization states of the same element. To determine the ionization parameter, we use the [OIII]$\lambda5007$/[OII]$\lambda3726, \lambda3729$ (O32) diagnostic.
\par
KD02 and KK04 both presented theoretical calibrations for ionization parameter using the O32 diagnostic. However, the O32 diagnostic has a strong dependence on metallicity. Unlike the \rtt diagnostic, the O32 diagnostic is unambiguous in the sense that it is not double valued except at high metallicities ( Z > 2 $\textrm{Z}_\odot$). At lower metallicities, the polynomial fits to the theoretical relationship between ionization parameter and the [OIII]/[OII] line ratio monotonically increase across the valid ionization parameter range.
\subsection{Iteration}
\label{iter}
We determine the metallicity and ionization parameter simultaneously through an iterative process. We first use an initial metallicity estimate to constrain the \rtt diagnostic to the upper or lower metallicity branch. The [NII]$\lambda6583$/[OII]$\lambda3726, \lambda3729$ (N2O2) diagnostic has very little dependence on ionization parameter (but is strongly affected by attenuation), and we use this diagnostic ratio for our initial metallicity estimate. For spaxels with N2O2$<-1.2$, we place the spaxel on the lower \rtt branch and assume a metallicity of $12+\log({\rm O/H}) = 8.2$ as the starting iteration point. For N2O2$>-1.2$, we use the upper \rtt branch and assume a metallicity of $12+\log({\rm O/H}) = 8.7$.
\par
Once an initial metallicity estimate has been determined, we use this value in the first estimate of the ionization parameter using the O32 diagnostic. This first ionization parameter estimate is then used to improve our metallicity estimate through the \rtt diagnostic. We continue iterating between the \rtt and O32 diagnostics until the metallicity and ionization parameter converge. We consider the metallicity and ionization parameter converged if the difference between iterations in the metallicity estimate is less than 0.1 dex and the ionization parameter estimate is within 0.01 dex. We require this tolerance to be achieved for all spaxels used during analysis. 
\par
The rate at which the metallicity and ionization parameter converge is usually proportional to the S/N ratio. Spaxels with a S/N > 5 in the used emission lines generally converge in $\sim$ 3 iterations while lower S/N spaxels sometimes require 20+ iterations. We impose an upper limit of 20 iterations to remove any non-converging spaxels from the maps. Spaxels that have not converged are discarded from the metallicity and ionization parameter maps.

\subsection{Error Propagation}
\label{errorprop}
The iterative method used to calculate the metallicity and ionization parameter makes it difficult to analytically propagate the error. To propagate line flux errors produced by LZIFU through to the metallicity and ionization parameter, we simulate 1000 maps for all emission lines used in the calculation. The maps are created such that the fluxes are Gaussian distributed with the LZIFU standard deviation for that emission line. 
\par
Using the simulated line maps, metallicity and ionization parameter maps are created using the iterative process described in Section \ref{iter}. The non-linearity of the resulting metallicity and ionization parameter diagnostics means that the metallicity and ionization parameter distributions are not necessarily Gaussian. To represent the spread of metallicity and ionization parameter, we determine the distance from the best-fit value to the $16^{\textrm{th}}$ and $84^{\textrm{th}}$ percentiles and calculate the average. This provides us with a measure of the uncertainty of the metallicity and ionization parameter maps which are then propagated to the gradient errors.

\section{Metallicity Distribution}
\begin{landscape}
	\begin{figure}
	\centering
	\includegraphics[width=1.3\textheight]{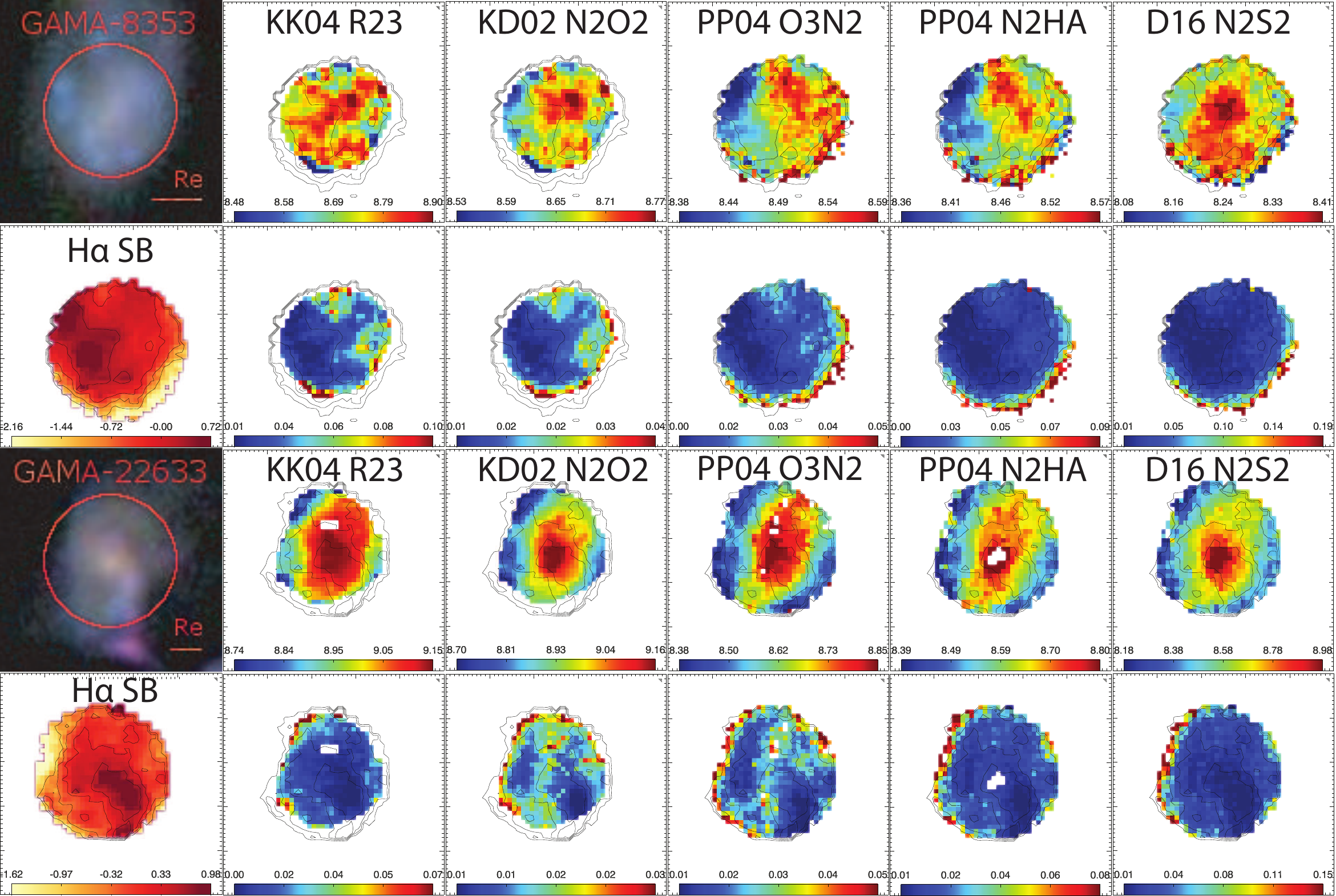}\label{fig:a}
	\caption{
Each galaxy is presented as a 2$\times$6 grid. The top row of each grid shows the various metallicity maps with their corresponding error maps beneath them. Note that scale bars have been varied between different maps and galaxies in order to provide the best metallicity resolution possible. \textit{Column 1:} SDSS composite image obtained from DR10. The red circles represent the 14.7'' aperture of the SAMI hexabundle and the scale bar shows the effective radius of the galaxy obtained from GAMA R band. Below this we show the H$\alpha$ emission line map. We choose the H$\alpha$ emission line map because we believe it provides the best representation of the galaxy structure and morphology. We overplot the H$\alpha$ contours onto each metallicity map to provide a point of reference when comparing metallicity diagnostics. \textit{Column 2:} KK04 metallicity determined from the \rtt line ratio. \textit{Column 3:} KD02 metallicity determined from the N2O2 line ratio. \textit{Column 4:} PP04 metallicity determined from the O3N2 line ratio. \textit{Column 5:} PP04 metallicity determined from the N2HA line ratio. \textit{Column 6:} D16 metallicity determined from the N2S2 line ratio. All metallicity maps are measured in units of 12 + log(O/H).
	}
	\label{metmapgrid}
	\end{figure}
\end{landscape}

We calculate metallicity and ionization parameter maps with their corresponding errors for our sample of 25 SAMI galaxies. In Figure \ref{metmapgrid} we show two examples of the metallicity maps using different metallicity diagnostics accompanied by their error maps. In addition to the metallicity maps, we also show the SDSS 3 colour image (gri) of the galaxy with the SAMI field of view and its effective radius. The metallicity maps for the other 23 galaxies are presented in the Appendix. The majority of galaxies in our sample have metallicities in the range $8.5 <$12$+$log(O/H)$<9.3$ in the radial ranges probed using the KK04 metallicity diagnostics. The mass-metallicity relation presented in \citet{kewley08} shows that the nuclear metallicities for SDSS galaxies range between $8.7<$12$+$log(O/H)$<9.05$ for a mass range between $9.0 <\log{M_{*}/M_{\odot}}<10.5$. This is consistent with the metallicites within our sample for the same mass range, given that the SDSS fibre samples $\sim20\%$ of the galaxies' B-band light \citep{kewley05}.
\par
\begin{landscape}
	\begin{figure}
	\centering
	\includegraphics[width=1.3\textheight]{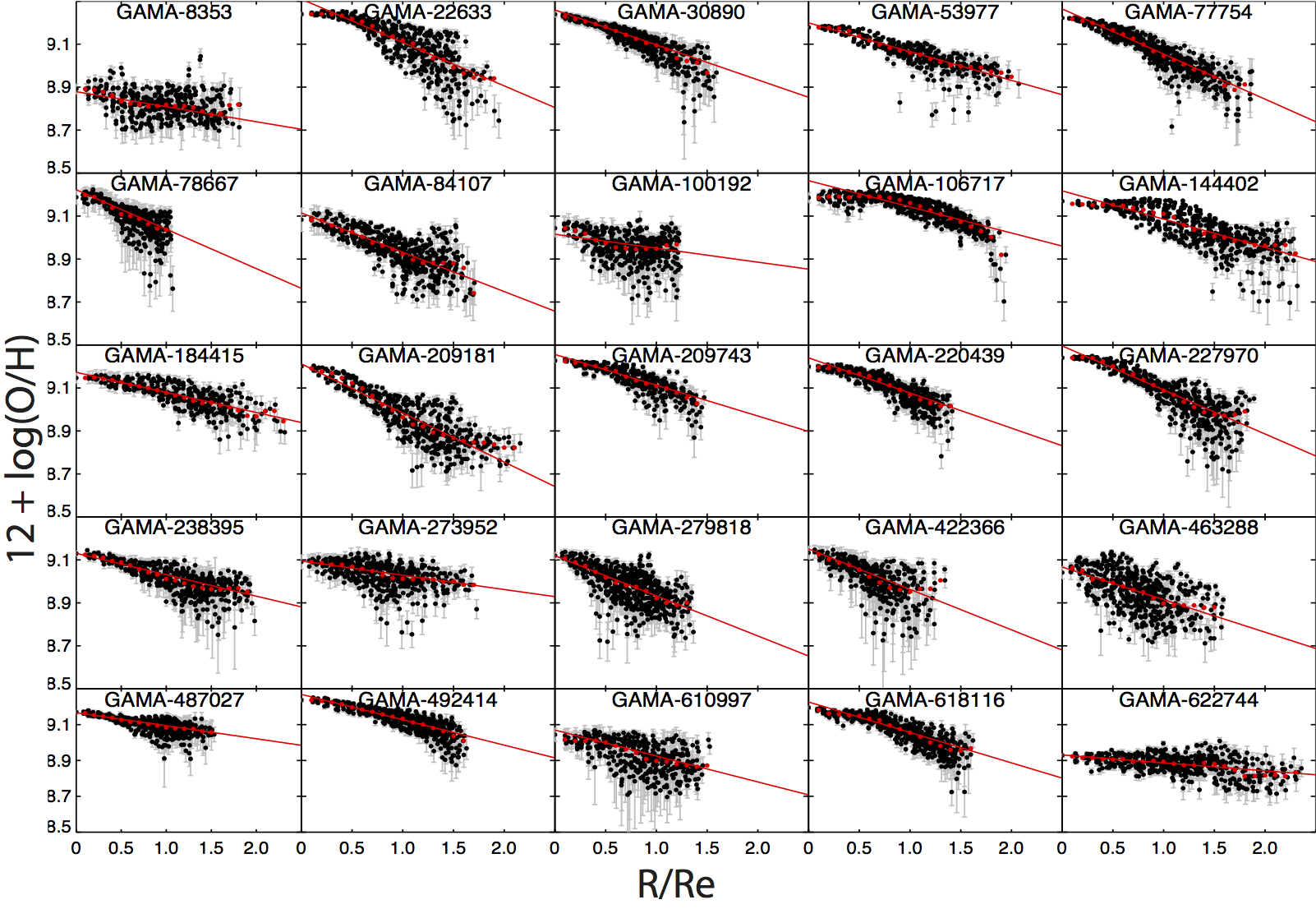}\label{fig:a}
	\caption{
		KK04 \rtt metallicity gradients used in our analysis. We show 1$\sigma$ error bars for each spaxel, determined from the method described in Section \ref{errorprop}. The best linear fit to the metallicity gradient is given as a red line. The median metallicity in bins of 0.1 R/$R_{e}$ are filled red circles. The results are summarised in Table \ref{mettable}.
	}
	\label{metgradgrid}
	\end{figure}
\end{landscape}

The gas-phase metallicity increases over time. For the inside-out model of galaxy formation, we expect isolated galaxies to have strong negative metallicity gradients \citep{pagel81,edmunds84,vilchez88,vila-costas92,zaritsky94}. However, in interacting galaxies, the turbulent gas caused by the tidal forces stretches and flattens this metallicity gradient \citep{torrey12}. We find that in our sample, the majority of galaxies possess strong metallicity gradients (18/25), as expected for relatively isolated and undisturbed galaxies. We show these normalized metallicity gradients in Figure \ref{metgradgrid} and provide a table of each linear fit in Table \ref{mettable}. 
\par
\begin{landscape}
	\begin{figure}
	\centering
	\includegraphics[width=1.3\textheight]{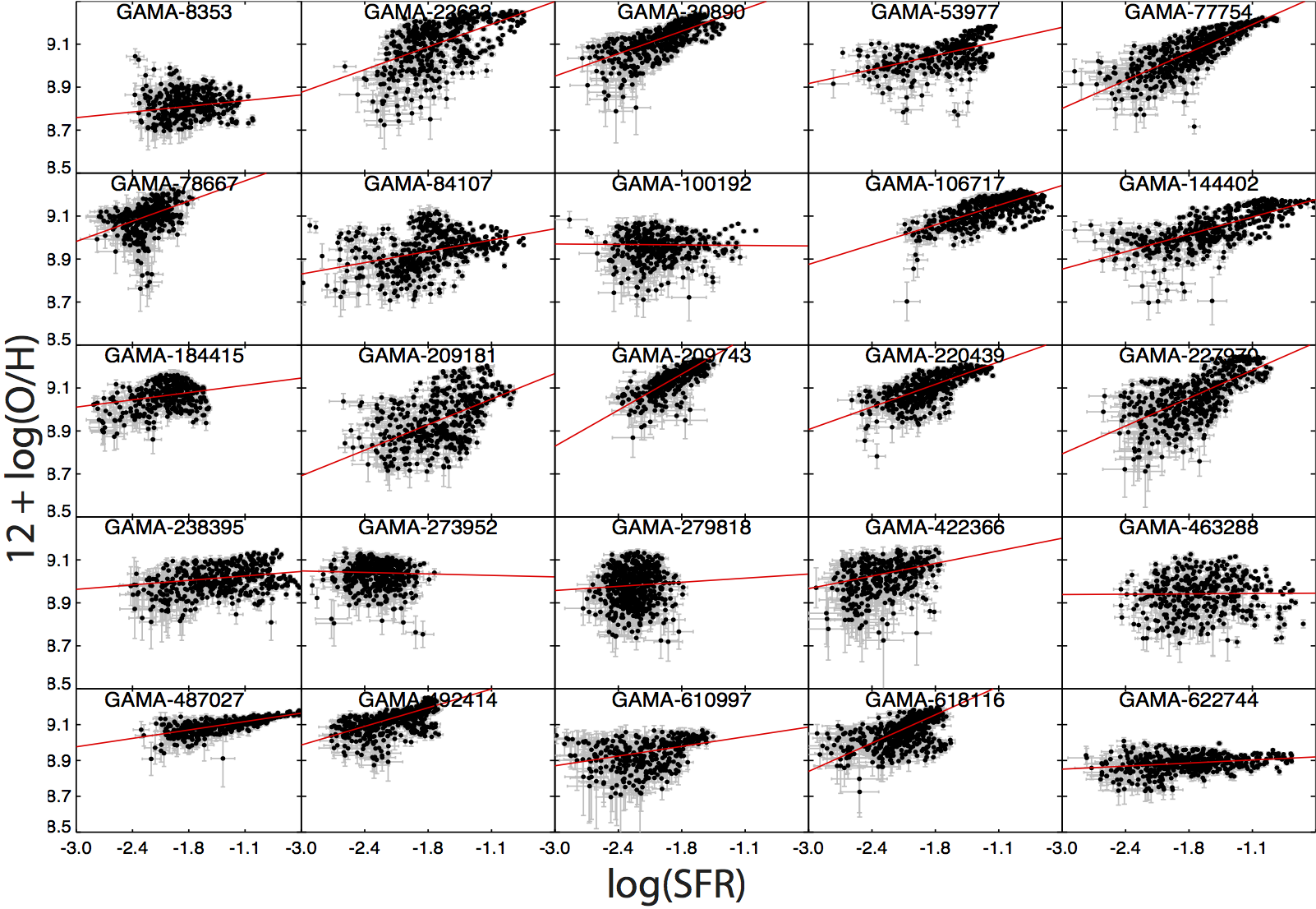}\label{fig:a}
	\caption{
		KK04 \rtt metallicity as a function of SFR surface density. We obtain maps of SFR surface density in units of M$_{\odot}$/year/kpc$^2$ from \citet{medling18}. We show the best linear fit as a red line and summarise the results in the Table \ref{R23SFR}.
	}
	\label{SFRmetgradgrid}
	\end{figure}
\end{landscape}

For several galaxies we also find a strong positive correlation between metallicity and SFR surface density as shown in Figure \ref{SFRmetgradgrid}. This is consistent with several recent studies using SAMI data have shown SFR surface density gradients exist in the SAMI sample \citep{schaefer17,medling18}.
\par
\citet{sanchez-menguiano16b,ho17} showed that significant azimuthal variations exist in the metallicity distribution of NGC6754 and NGC1365 respectively. However, we split each galaxy into quadrants and find little evidence of significant changes in the metallicity gradient, suggesting that at the spatial resolution of SAMI, spatial smoothing is sufficient to remove any trace of azimuthal variations, leaving only the radial gradients we observe. A resolution of at least 200pc/PSF is needed to observe these azimuthal variations \citep{sanchez-menguiano16b}. With a median redshift of $z=0.028$, an average seeing of $2.16\arcsec$ combined with the $0.8\arcsec$ smoothing to remove DAR, our galaxy sample has median resolution elements of 1.3kpc/PSF, much coarser than the minimum requirement found in \citet{sanchez-menguiano16b}. While statistically significant azimuthal variations are absent, there is evidence of clumpy substructure in several metallicity maps (eg. GAMA-8353 and GAMA-106717). 

\clearpage

\subsection{Metallicity Gradients}
While there is little azimuthal variation in the sample, there are clear radial gradients across our sample. The smoothness of these metallicity maps means that we are able to use a simple linear fit to the metallicity ($12+\log(\rm O/H)$) as a function of radius. \citet{sanchez-menguiano18} showed that broken linear fits can also be used to describe the metallicity gradients of MUSE galaxies. Broken linear fits allow for the fitting of steepening or flattening metallicity gradients, resulting in a more robust fit for metallicity gradients which vary with radii. For this study, we use single linear fits to the galaxy metallicity gradients.
\par
We determine the radial distance of each pixel from the centre taking into account the ellipticity and position angle of the object. We also normalise the radius by the size of its disk using its effective radius (Re) in the R band measured using GALFIT \citep{peng02} by the GAMA survey \cite{kelvin12}. This removes the size dependence that the metallicity gradient has when measured on a physical scale \citep{sanchez14,ho15}.
\par
We use the robust line fitting routine LTS$\_$LINEFIT \citep{cappellari13} to fit a linear trend to the data. We choose LTS$\_$LINEFIT for its ability to identify and separate outliers from the input data as well as provide standard errors to the output fit parameters. To calculate the standard errors on the output fit parameters, we provide LTS$\_$LINEFIT with the metallicity errors calculated from method described in Section \ref{errorprop}. We show the radial metallicity gradients of our sample in Figure \ref{metgradgrid} along with the best linear fit and radially binned median points.
\par
The Pearson correlation coefficient (PCC) is a measure of the presence of a linear trend. A magnitude of greater than 0.6 is usually accepted as a strong indication of a linear trend. The majority of radial metallicity gradients determined by LTS$\_$LINEFIT show strong a strong trend (PCC magnitude > 0.6), with four galaxies presenting with very strong Pearson correlation coefficient (PCC magnitude > 0.8). 
\par
\begin{figure}
\centering

\begin{tabular}{c}
  \subfloat{\includegraphics[width=\linewidth]{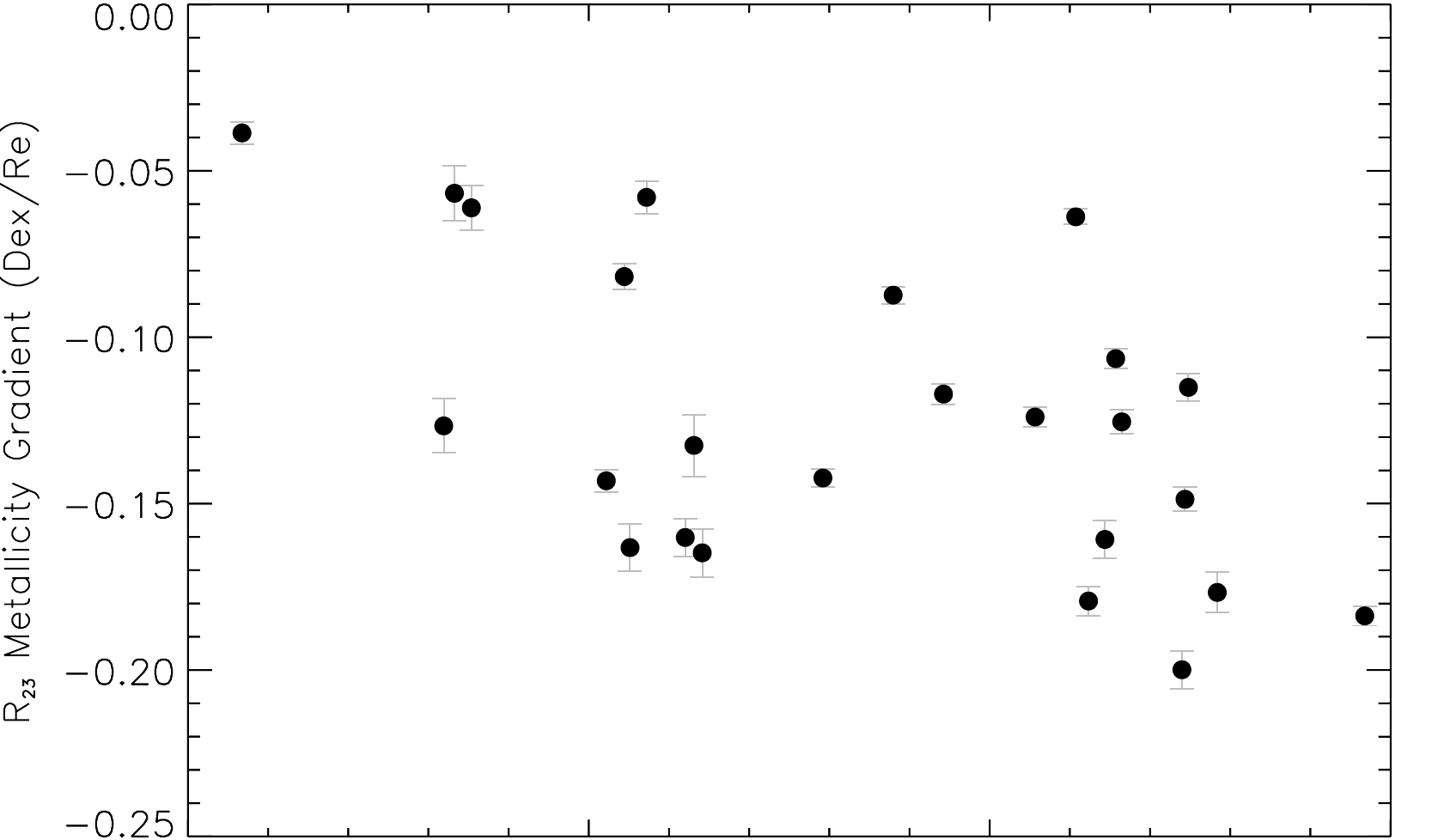}\label{fig:a}}\\[-1.5em]

  \subfloat{\includegraphics[width=\linewidth]{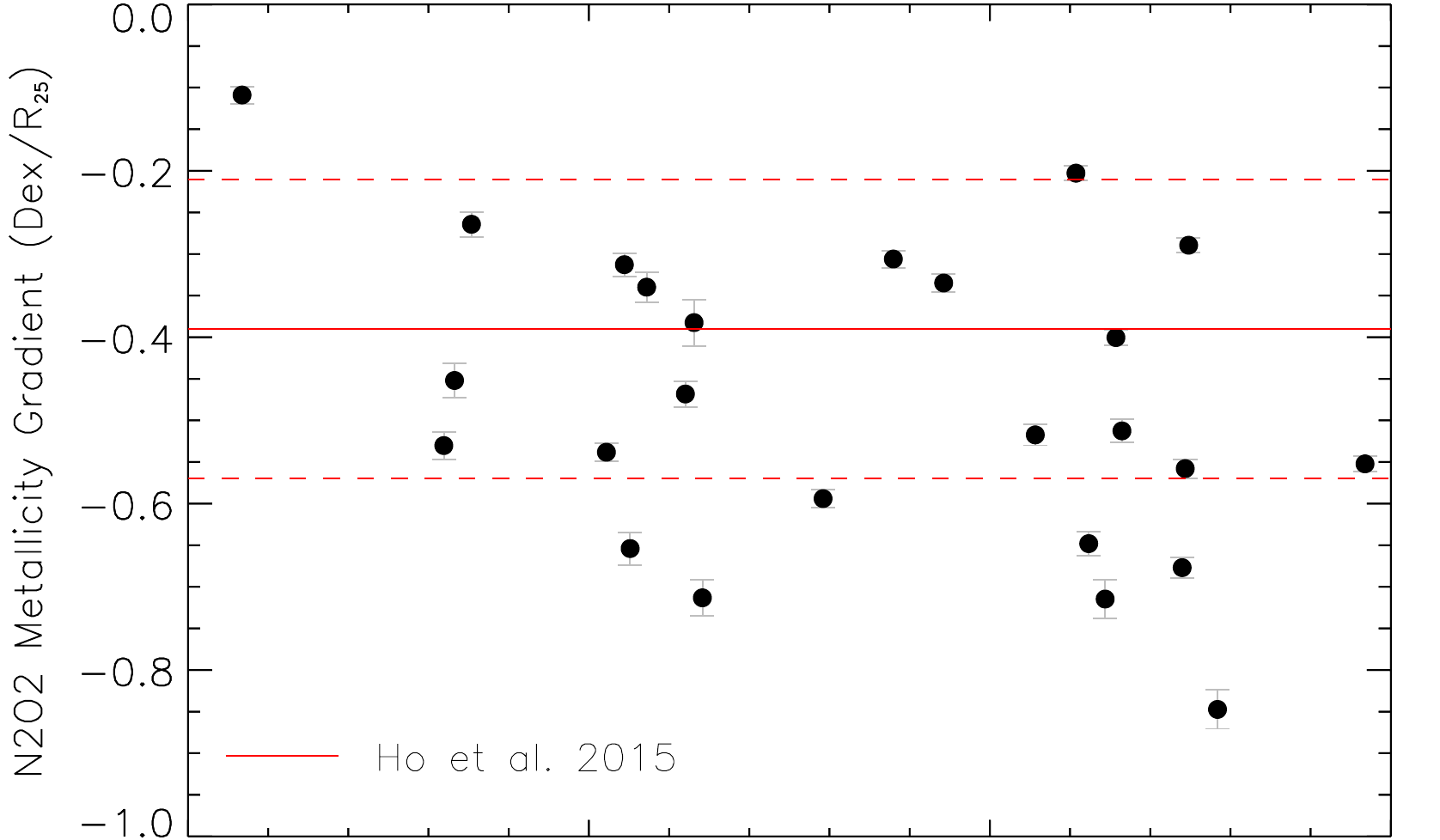}\label{fig:a}}\\[-1.5em]

  \subfloat{\includegraphics[width=\linewidth]{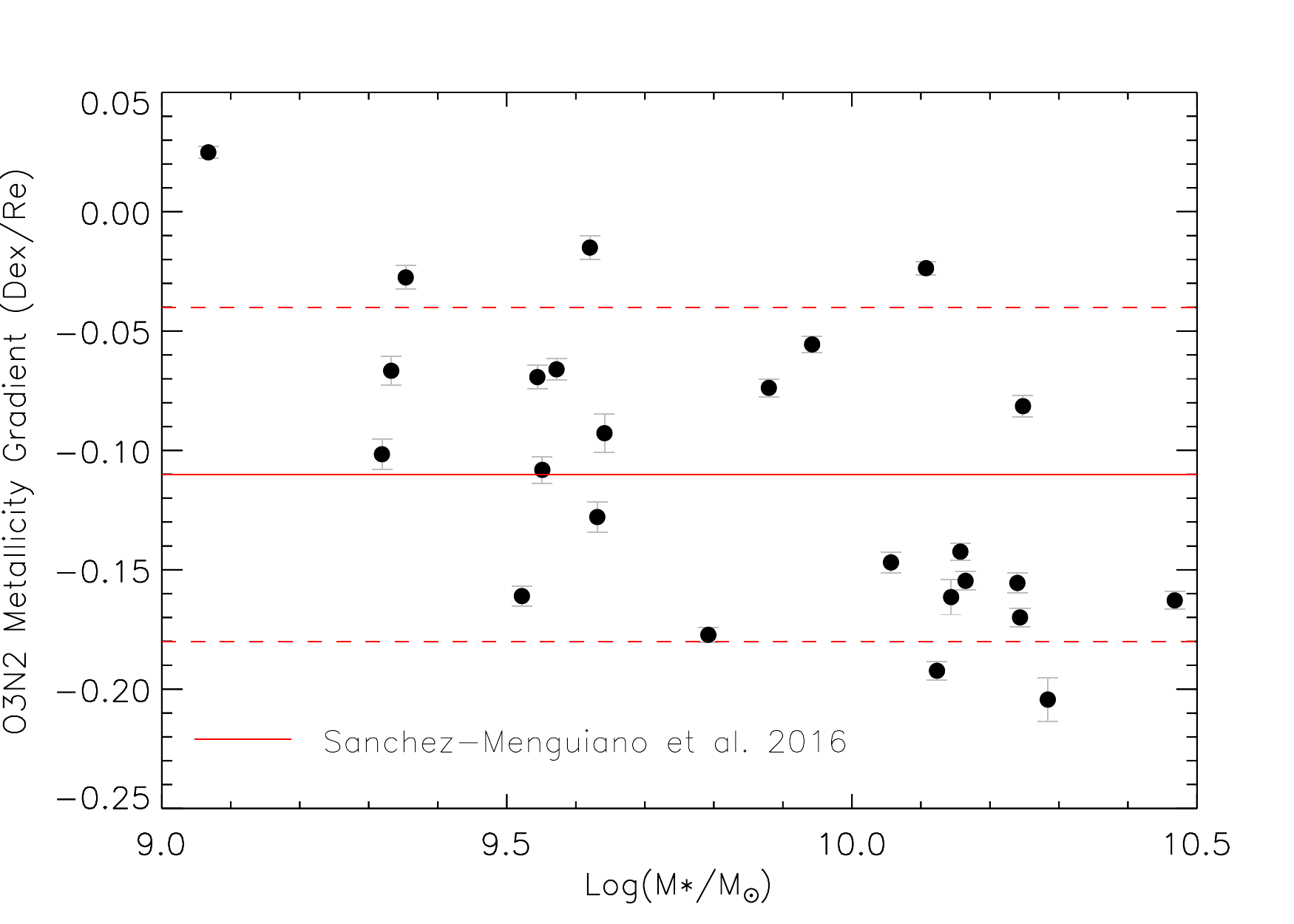}\label{fig:a}}\\

\end{tabular}
\caption{
Normalised metallicity gradients as a function of mass using 3 common metallicity diagnostics. For the KD02 and PP04 metallicity diagnostics, we compare the results presented in \citet{ho15} and \citet{sanchez-menguiano16}. The solid red line show the mean metallicity gradient with 1$\sigma$ scatter shown as dotted red lines. 
}
\label{metgrad}

\end{figure}

Figure \ref{metgrad} shows the normalised metallicity gradients of galaxies against their stellar masses. Within our mass range of $9.0 < \log(M_{*}/M_{\odot}) < 10.5$, the normalised metallicity gradients range from -0.20 to -0.03 dex/Re. There appears to be a slight correlation with steeper metallicity gradients occurring at higher masses. We fit the relationship with a linear trend and find a slope of $-0.065\pm0.021$ dex/Re/$\log(M_{*}/M_{\odot})$ with a PCC of -0.54. \citet{belfiore17} finds a similar trend with steeper metallicity gradients occurring in more massive galaxies in the mass range $9.0 < \log(M_{*}/M_{\odot}) < 10.5$.
\par
Estimating the error on the PCC through bootstrapping analysis, we find PCC$=-0.54\pm0.06$ for the relationship between stellar mass and metallicity gradients. This indicates that there exists a weak negative linear trend between stellar mass and metallicity gradients for galaxies in the mass range $9.0 < \log(M_{*}/M_{\odot}) < 10.5$. This disagrees with previous studies by \citet{sanchez12,sanchez14,ho15,sanchez-menguiano16} who found no variation in radial metallicity gradients in their sample when normalised with either R$_{25}$ or $R_{e}$.
\subsection{Mass-Metallicity Relation}
\begin{figure}
\centering

\begin{tabular}{c}
  \subfloat{\includegraphics[width=\linewidth]{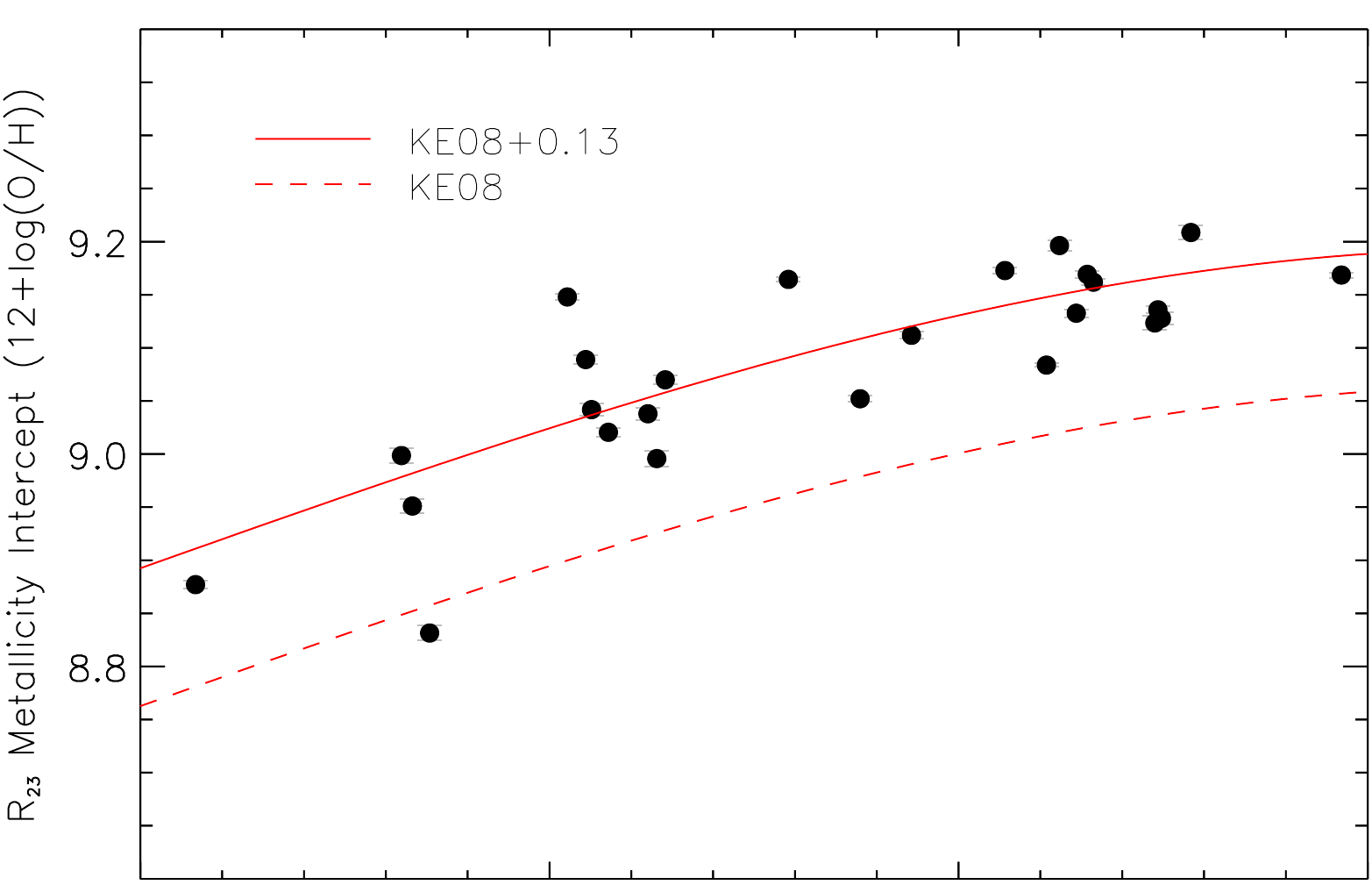}\label{fig:a}}\\[-1.5em]

  \subfloat{\includegraphics[width=\linewidth]{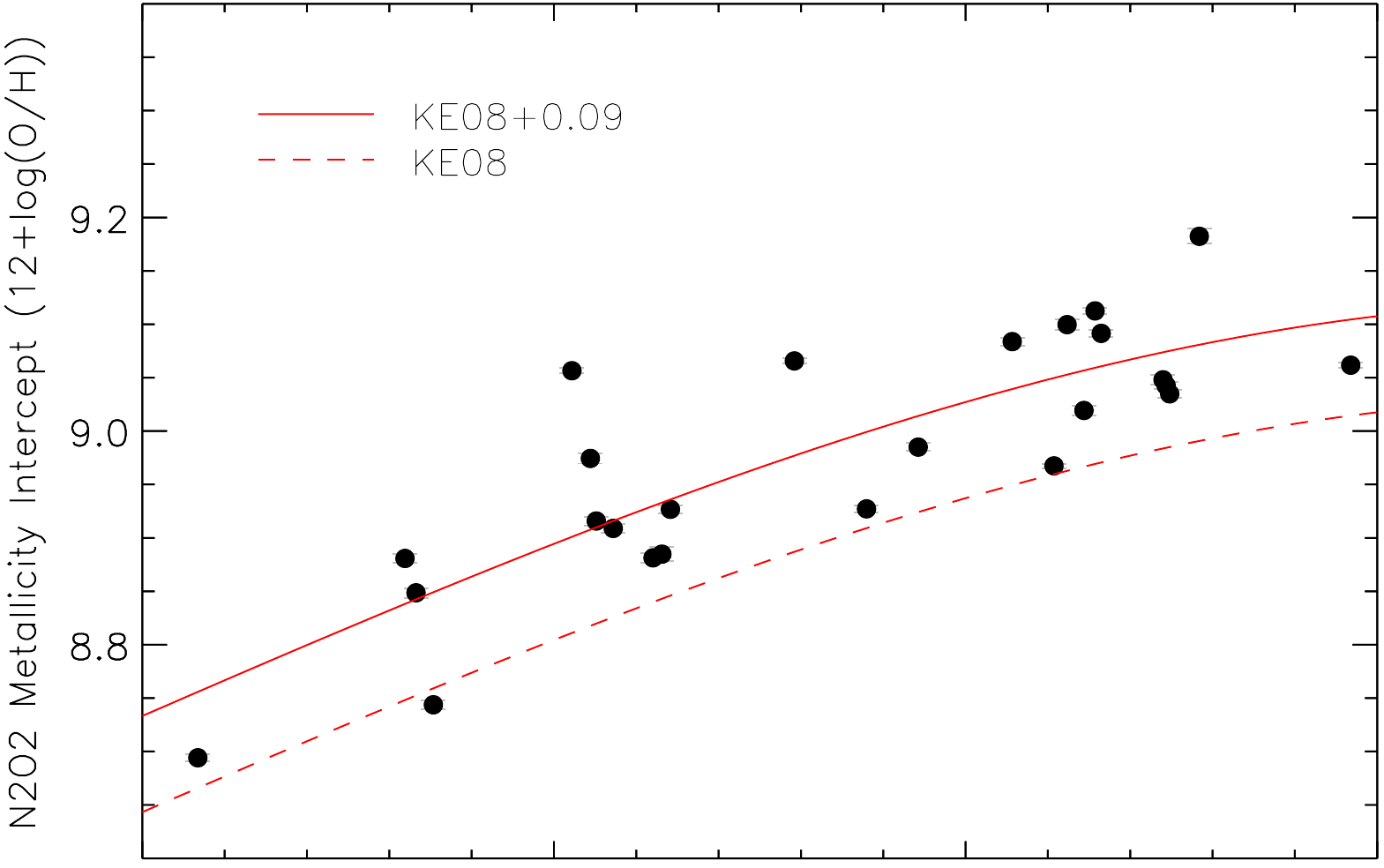}\label{fig:a}}\\[-1.5em]

  \subfloat{\includegraphics[width=\linewidth]{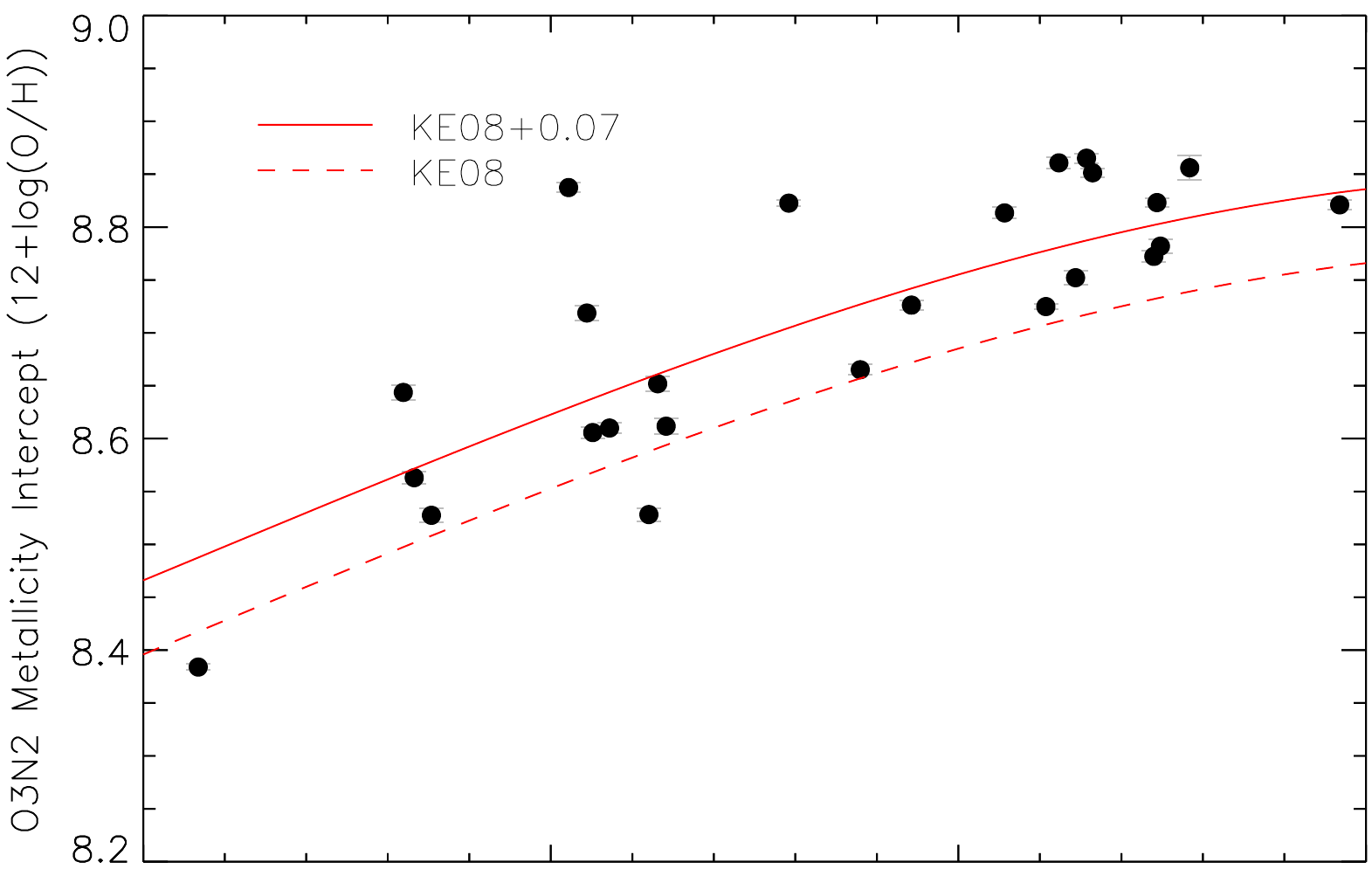}\label{fig:a}}\\[-1.5em]

  \subfloat{\includegraphics[width=1.06\linewidth]{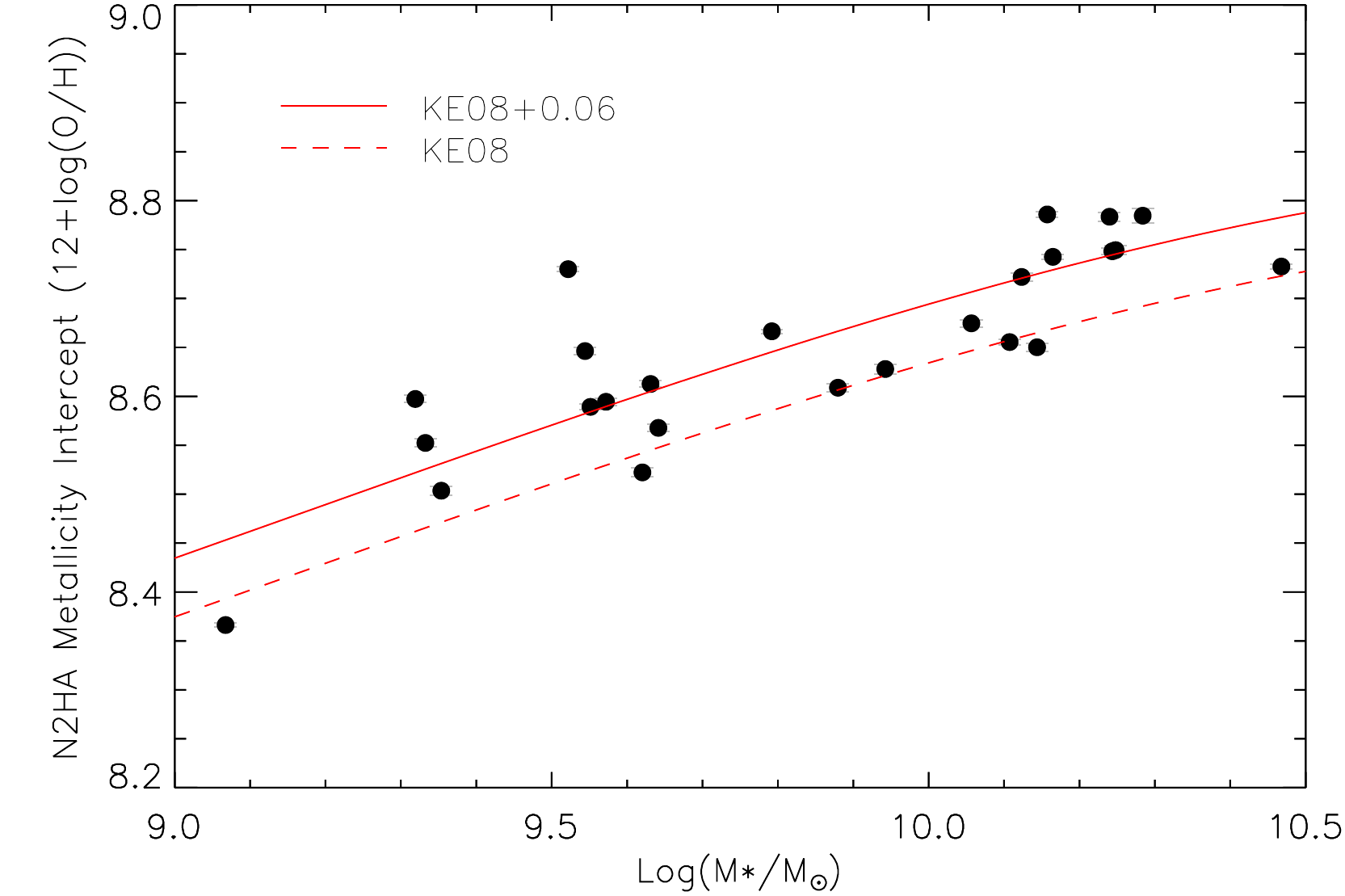}\label{fig:a}}\\

\end{tabular}

\caption{
Metallicity intercepts as a function of mass for multiple metallicity diagnostics. We show the mass-metallicity relation for each diagnostic from \citet{kewley08} as the dotted red line and fit an offset shown as the solid red line. 
}
\label{metint}

\end{figure}

While the radial metallicity gradients appear to be weakly dependent on galaxy masses, across $9.0 < \log(M_{*}/M_{\odot}) < 10.5$ there still exists the global mass-metallicity relation. Figure \ref{metint} shows the correlation of the metallicity intercept with stellar mass for multiple metallicity diagnostics.
\par
\citet{kewley08} provides fits to the mass-metallicity relation for a range of different metallicity diagnostics. We plot the mass-metallicity fit for several metallicity diagnostics as the dotted red line on Figure \ref{metint}. There is a clear offset between the mass-metallicity fit and the metallicity intercepts caused by using the central interpolated metallicities rather than aperture metallicities. We fit these offsets using MPFIT \citep{markwardt09} and show the best least squares fit to the interpolated metallicities. Similar trends with the metallicity intercept were found in \citet{sanchez14} who also attributed it to the mass-metallicity relation.
\par

\renewcommand{\tabcolsep}{0.15cm}
\begin{table}
\centering
\begin{tabular}{l*{5}{c}}
GAMA ID & Central Metallicity & Gradient & RMS & PCC\\
&	12+log(O/H)&	dex/Re& \\
\hline
008353&8.831$\pm$0.007&-0.061$\pm$0.007&0.081&-0.33\\
022633&9.209$\pm$0.006&-0.177$\pm$0.006&0.056&-0.71\\
030890&9.164$\pm$0.002&-0.142$\pm$0.003&0.034&-0.87\\
053977&9.112$\pm$0.003&-0.117$\pm$0.003	&0.027&-0.69\\
077754&9.169$\pm$0.002&-0.184$\pm$0.003&0.032&-0.89\\
078667&9.133$\pm$0.004&-0.161$\pm$0.006&0.037&-0.70\\
084107&9.038$\pm$0.006&-0.160$\pm$0.006&0.047&-0.65\\
100192&8.951$\pm$0.007&-0.057$\pm$0.008&0.061&-0.33\\
106717&9.169$\pm$0.003&-0.106$\pm$0.003&0.030&-0.72\\
144402&9.128$\pm$0.006&-0.115$\pm$0.004&0.049&-0.68\\
184415&9.089$\pm$0.004&-0.082$\pm$0.004&0.033&-0.76\\
209181&9.123$\pm$0.007&-0.200$\pm$0.006&0.073&-0.77\\
209743&9.162$\pm$0.003&-0.125$\pm$0.004&0.029&-0.84\\
220439&9.148$\pm$0.003&-0.143$\pm$0.003&0.027&-0.79\\
227970&9.196$\pm$0.005&-0.179$\pm$0.004&0.065&-0.75\\
238395&9.052$\pm$0.003&-0.087$\pm$0.003&0.038&-0.72\\
273952&9.020$\pm$0.004&-0.058$\pm$0.005&0.041&-0.44\\
279818&9.042$\pm$0.005&-0.163$\pm$0.007&0.073&-0.30\\
422366&9.070$\pm$0.004&-0.165$\pm$0.007&0.067&-0.61\\
463288&8.996$\pm$0.008&-0.133$\pm$0.009&0.080&-0.48\\
487027&9.084$\pm$0.002&-0.064$\pm$0.002&0.025&-0.65\\
492414&9.173$\pm$0.003&-0.124$\pm$0.003&0.032&-0.84\\
610997&8.999$\pm$0.007&-0.127$\pm$0.008&0.082&-0.51\\
618116&9.136$\pm$0.003&-0.149$\pm$0.004&0.036&-0.78\\
622744&8.877$\pm$0.004&-0.039$\pm$0.003&0.048&-0.47\\
\end{tabular}
\caption{List of metallicity gradients and intercepts with their 1$\sigma$ uncertainties, root mean square (RMS) scatter and Pearson correlation coefficient (PCC) values.}
\label{mettable}
\end{table}
 
\section{Ionization Parameter Distribution}
\subsection{Ionization Parameter Gradients}
\begin{figure}
\begin{centering}
\includegraphics[width=\linewidth]{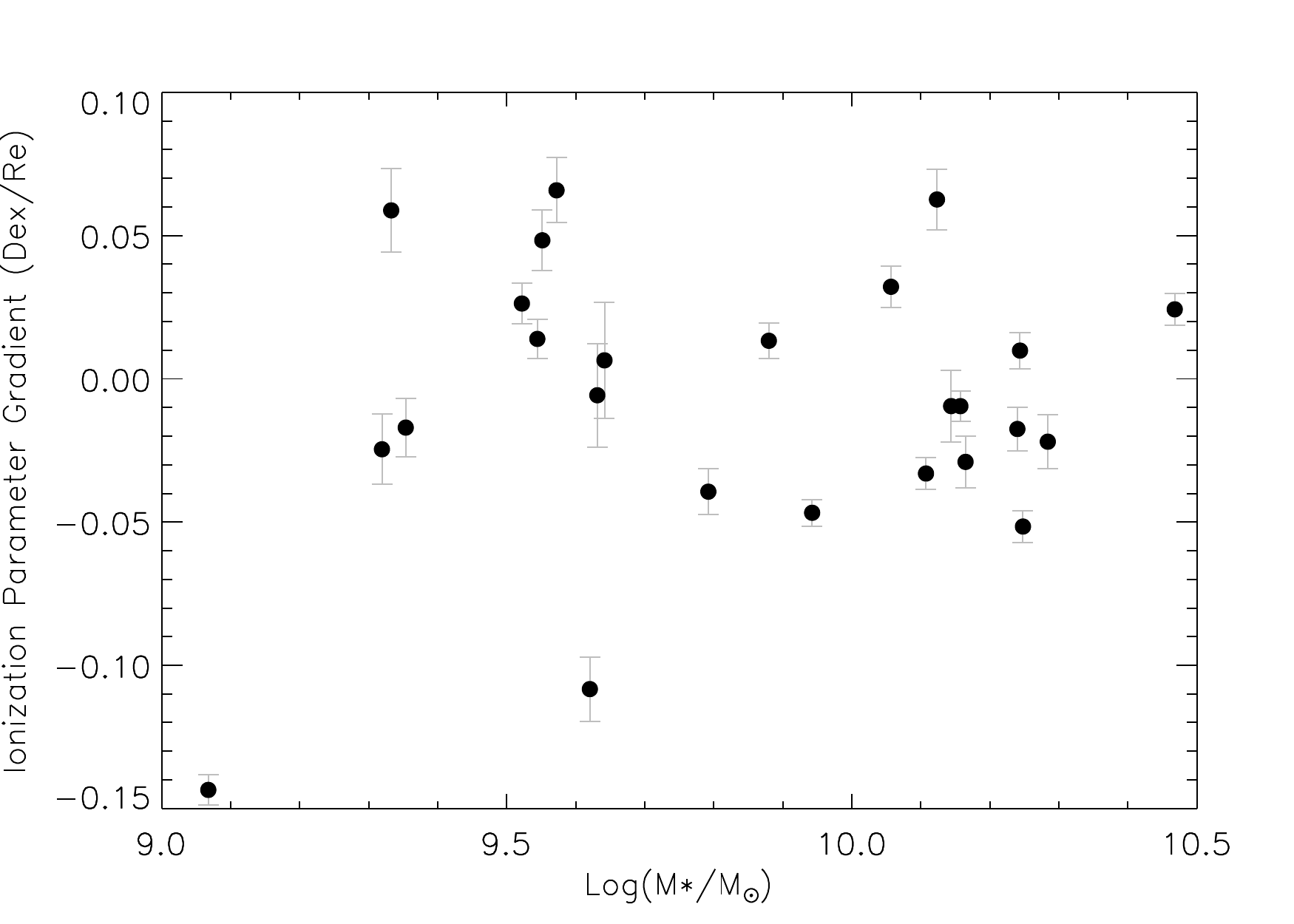}
\end{centering}
\caption{
Normalized ionization parameter gradients using the KK04 O32 diagnostic as a function of stellar mass. We find no significant variation in the ionization parameter gradient as a function of galaxy mass. 
}
\label{iongrad}
\end{figure}

\begin{landscape}
	\begin{figure}
	\centering
	\includegraphics[width=1.3\textheight]{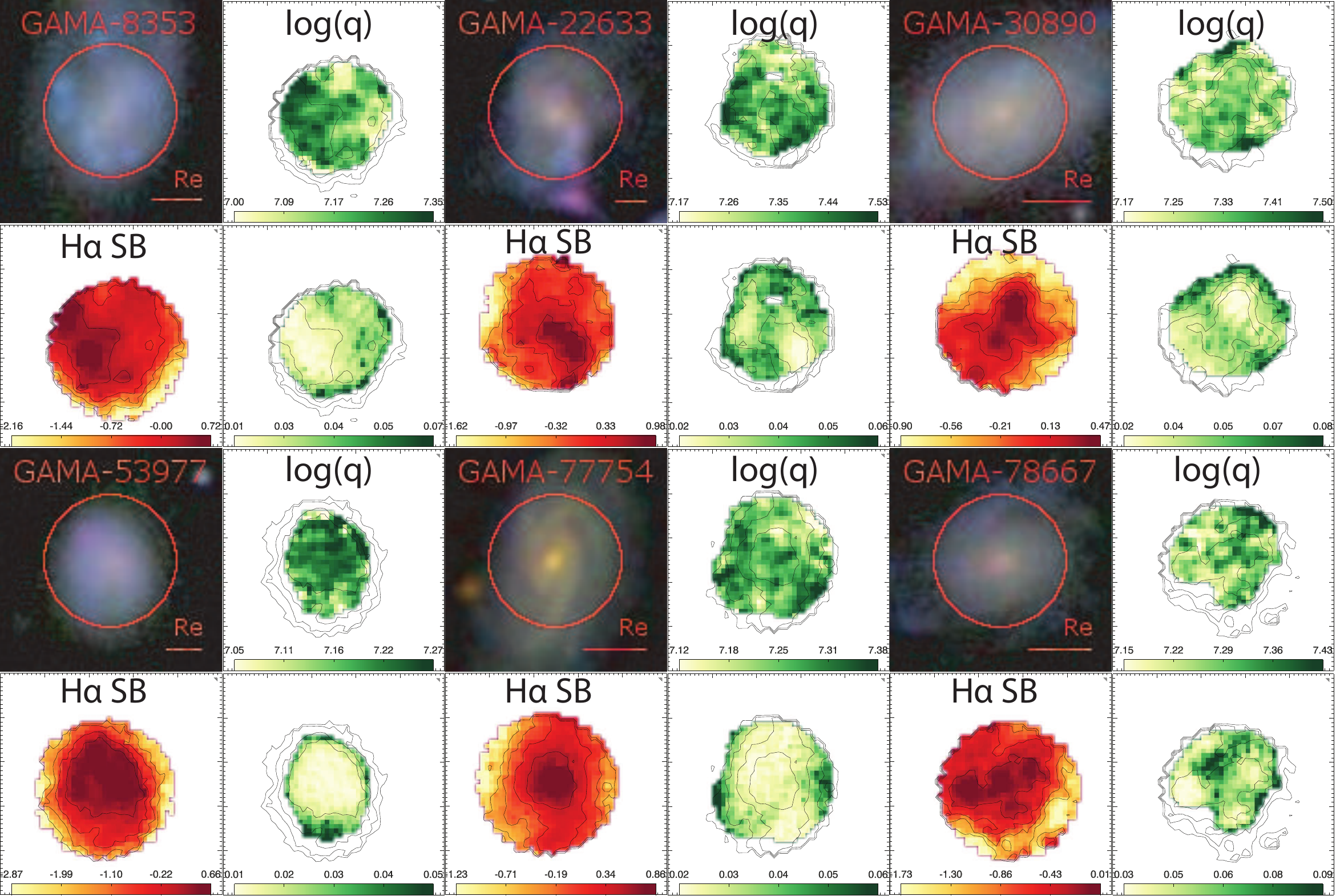}\label{fig:a}
	\caption{
Each galaxy is presented with a 2$\times$2 grid containing the results of our work. The top left image of each grid contains the same SDSS image as Figure \ref{metmapgrid}. The H$\alpha$ emission line map with contours below the SDSS image is also identical to Figure \ref{metmapgrid}. To the right of each SDSS image is the ionization parameter map in units of $\log$(cm/s) with overplotted H$\alpha$ contours for comparison. Below each ionization parameter map is the associated error map as described in Section \ref{errorprop}. Note that scale bars have been varied between different maps and galaxies in order to provide the best resolution possible
	}
	\label{ionmapgrid}
\end{figure}
\end{landscape}

\begin{landscape}
	\begin{figure}
	\centering
	\includegraphics[width=1.3\textheight]{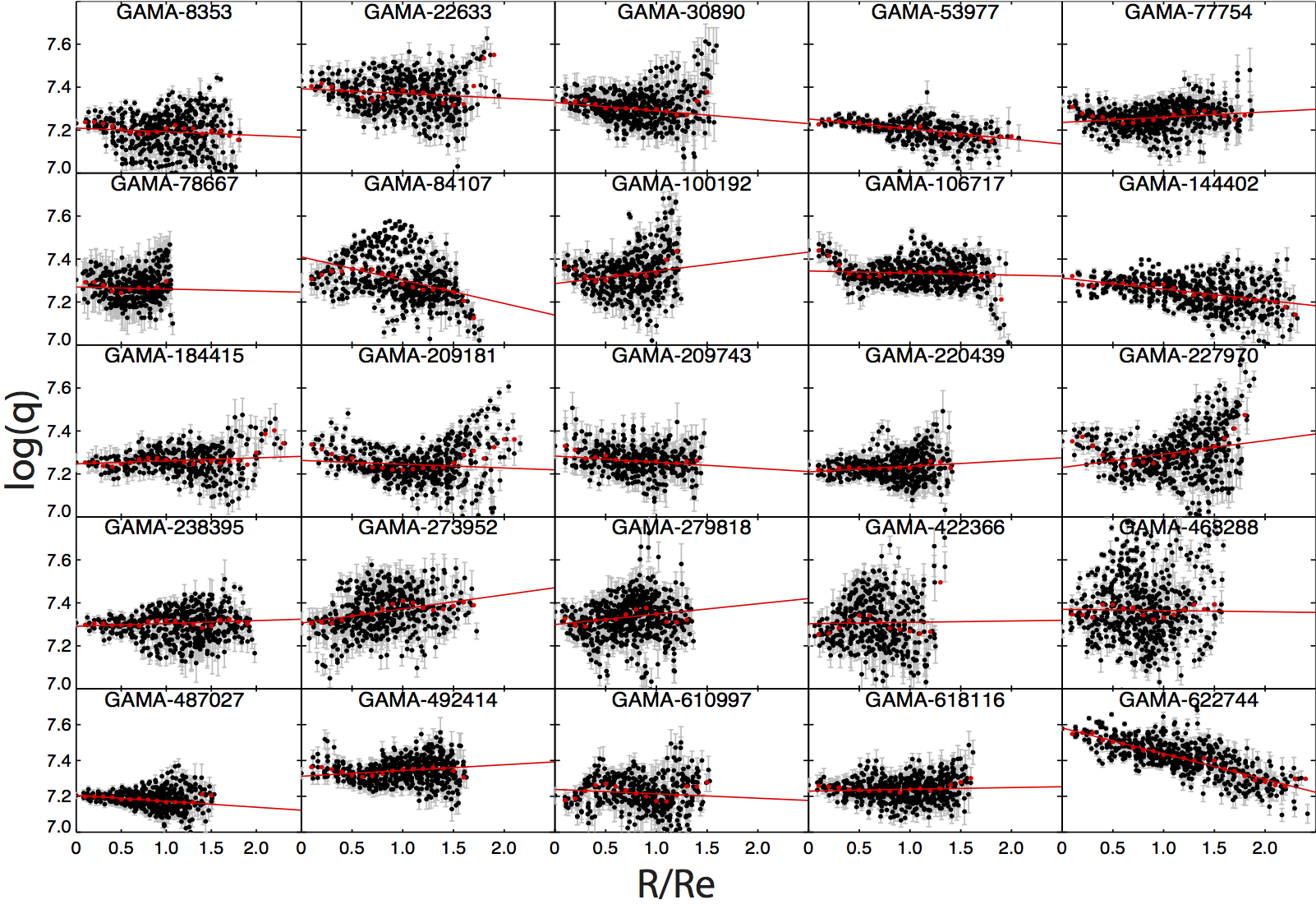}\label{fig:a}
	\caption{
	Same as Figure \ref{metgradgrid} for ionization parameter. The results are summarised in Table \ref{iontable}.
	}

\label{iongradgrid}
\end{figure}
\end{landscape}

In contrast to the metallicity maps, the ionization parameter maps (Figure \ref{ionmapgrid}) show no clear radial or azimuthal trends. Instead we see a range of different distributions ranging from weak gradients, flat maps and clumpy distributions. The majority of galaxies tend to have ionization parameters in the range $7.0 < \log(q$[cm/s]$) < 7.8$. We measure the radial ionization parameter gradients of the galaxies using robust line fits in the same way as the metallicity gradients. The ionization parameter radial gradients are presented in Figure \ref{iongradgrid} and compared to stellar mass in Figure \ref{iongrad}. All galaxies except three have a PCC magnitude of less than 0.4, indicating very weak significance of these linear fits. GAMA-622744 appears to be the only galaxy with a significant ionization parameter gradient (PCC magnitude = 0.73).
\par
\citet{kaplan16} found significant ionization parameter gradients in their sample of 8 galaxies using VENGA data. The galaxies in their sample were chosen to have significant and highly resolved bulges. \citet{kaplan16} used the same O32 ionization parameter diagnostics from KK04 as we do, but use one iteration rather than a convergence condition when calculating ionization parameter. Both methods provide them with similar results. The distribution of ionization parameter in their maps follows the distribution of SFR surface density in many of their galaxies and show strong radial gradients.
\subsection{Ionization Parameter and Galaxy Properties}
\begin{landscape}
	\begin{figure}
	\centering
	\includegraphics[width=1.3\textheight]{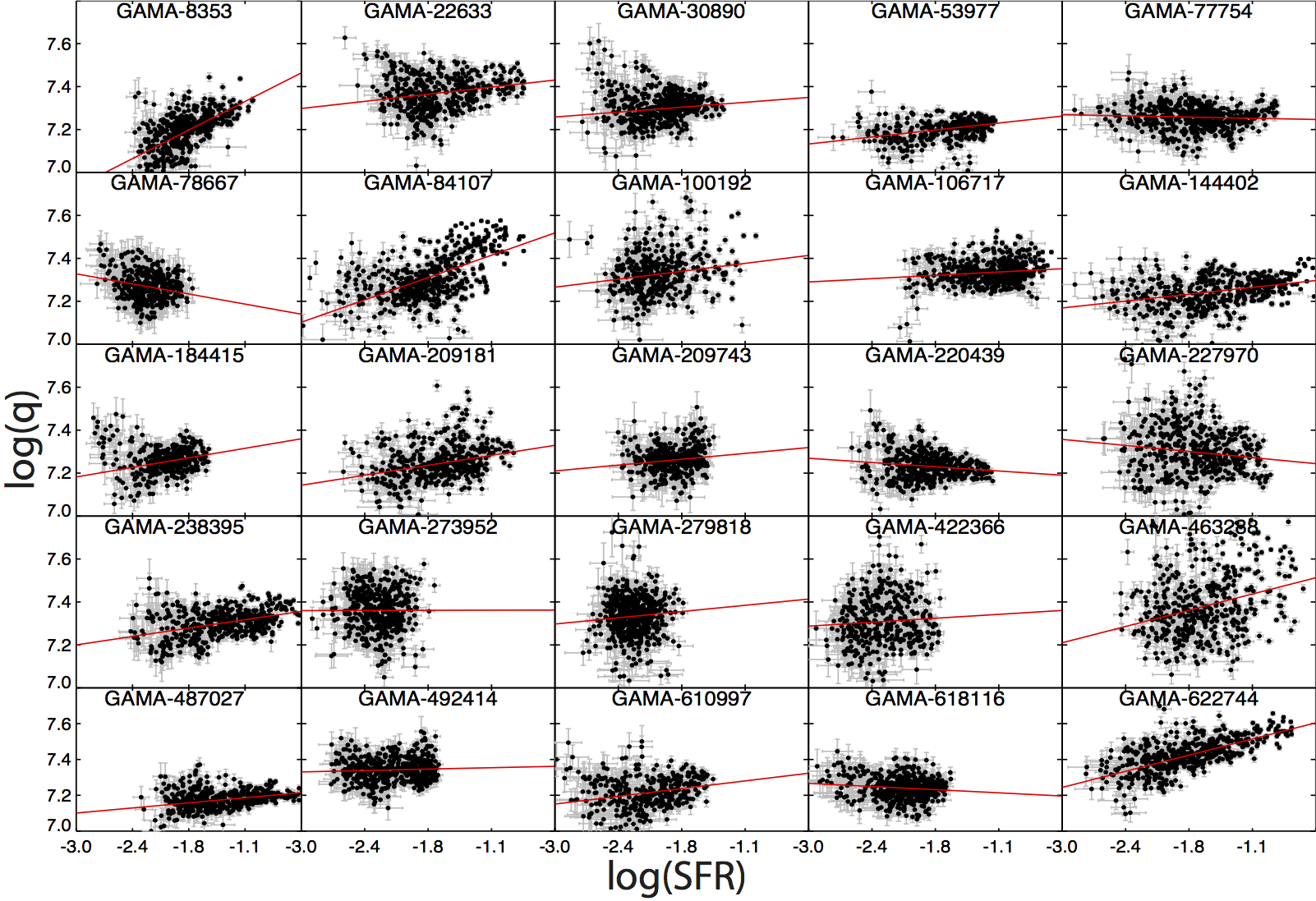}\label{fig:a}
	\caption{
		Same as Figure \ref{SFRmetgradgrid} for ionization parameter. We summarise the results in the Table \ref{R23QSFR}.
	}
	
	\label{SFRiongradgrid}
\end{figure}
\end{landscape}

While we see no significant radial or azimuthal trends in the ionization parameter for most of our sample, GAMA-8353 and GAMA-22633, show patterns in $q$ that are suggestive of the spiral arm features seen in the associated 3-colour and H$\alpha$ maps in each galaxy. Such an association could indicate that the ionization parameter is larger in areas of high star formation, a trend seen by \citet{dopita14} in a sample of luminous infrared galaxies (LIRGs) above a threshold ionization parameter ($\log(q[$cm/s$])> 7.2-7.4$). \citet{dopita14} quantified this relation as $q$[cm/s]$ \propto \textrm{SFR}$[M$_{\odot}$/year/kpc$^2$]$^{0.34\pm0.08}$. Using SFR surface density maps created by \citet{medling18}, we find that 71$\%$ (17/24) of galaxies present a slight positive correlation between SFR surface density and ionization parameter (Figure \ref{SFRiongradgrid}). However the strength of these gradients is weak with only GAMA-622744 having a PCC magnitude of greater than 0.6.
\par
\begin{landscape}
	\begin{figure}
	\centering
	\includegraphics[width=1.3\textheight]{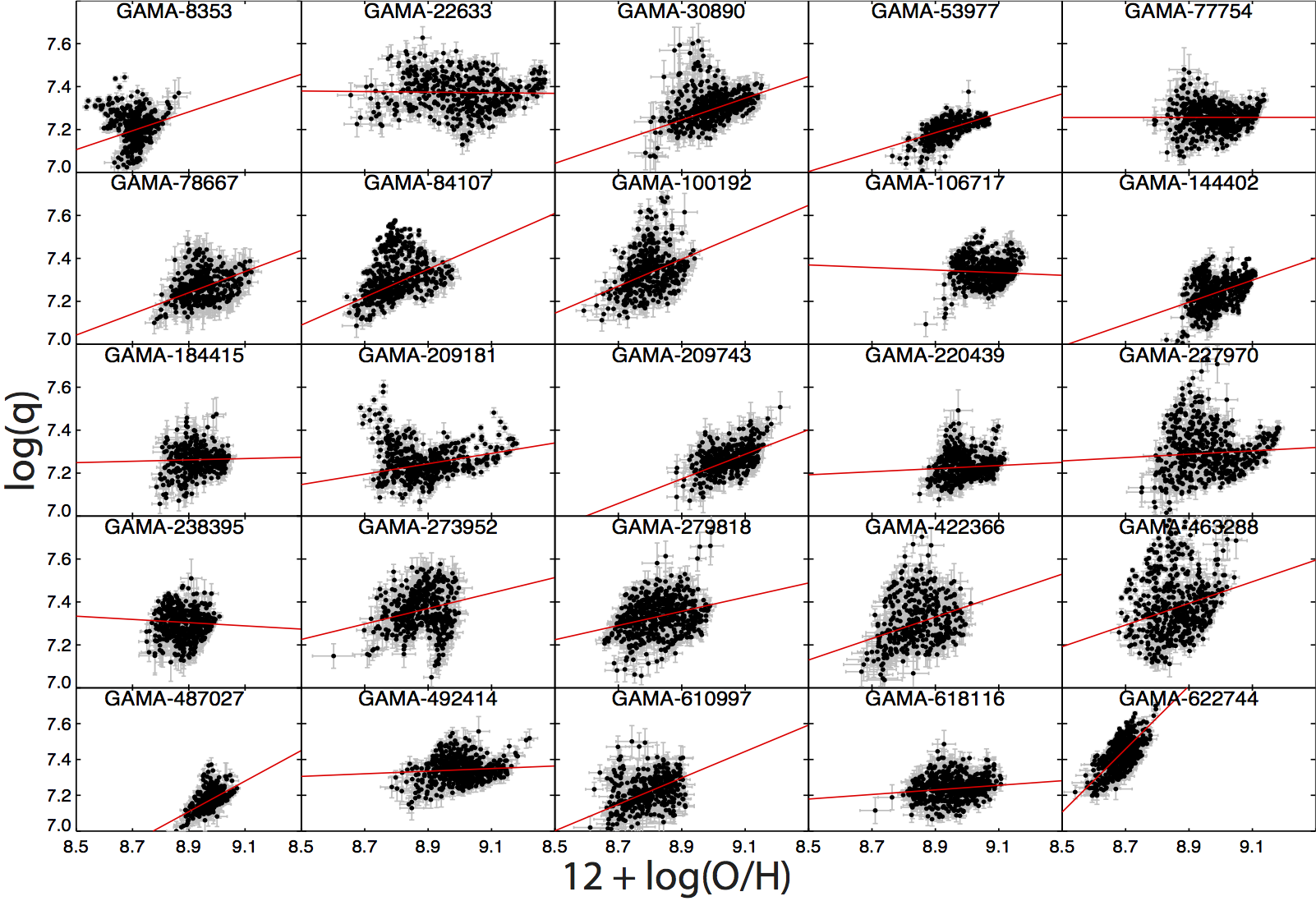}\label{fig:a}
	\caption{
		Relationship between the ionization parameter determined from the KK04 O32 diagnostic and the metallicity calculated from the KD02 N2O2 diagnostic. The best linear fit is given as a red line and we summarise the results in the Table \ref{R23QN2O2}.
	}
	
	\label{metiongrid}
\end{figure}
\end{landscape}

We also investigate how the ionization parameter varies with metallicity (Figure \ref{metiongrid}). We do this by plotting the KD02 metallicity determined from the N2O2 diagnostic against the KK04 ionization parameter measurements. We use the KD02 N2O2 metallicity diagnostic instead of the KK04 R23 metallicity diagnostic to try and exclude any possible dependencies between the two parameters caused by the iterative method used to calculate the ionization parameter. Again we find that only GAMA-622744 produces a significant Pearson correlation coefficient. The correlation between metallicity and ionization parameter for GAMA-622744 is likely driven by the fact that it is the only galaxy in our sample with a significant negative ionization parameter gradient and not necessarily because of an intrinsic correlation between metallicity and ionization parameter.
\par
\citet{dopita14} found a strong positive trend between the metallicity and ionization parameter, which is not seen in either this work or \citet{kaplan16}. \citet{dopita06} provide a theoretical relationship between gas-phase metallicity and ionization parameter, $q[$cm/s$] \propto \textrm{Z[O/H]}^{-0.8}$.

\clearpage

\renewcommand{\tabcolsep}{0.15cm}
\begin{table}
\centering
\begin{tabular}{l*{5}{c}}
GAMA ID & Central Ionization & Gradient & RMS & PCC\\
&	log($q$) &	dex/Re&  \\
\hline
008353&7.210$\pm$0.011&-0.017$\pm$0.010&0.086&-0.08\\
022633&7.392$\pm$0.010&-0.022$\pm$0.009&0.081&-0.12\\
030890&7.329$\pm$0.006&-0.039$\pm$0.008&0.059&0.00\\
053977&7.252$\pm$0.005&-0.047$\pm$0.005&0.038&-0.43\\
077754&7.236$\pm$0.005&0.024$\pm$0.006&0.051&0.18\\
078667&7.270$\pm$0.009&-0.010$\pm$0.013&0.060&0.08\\
084107&7.410$\pm$0.012&-0.108$\pm$0.011&0.091&-0.37\\
100192&7.286$\pm$0.012&0.059$\pm$0.015&0.084&0.24\\
106717&7.344$\pm$0.006&-0.010$\pm$0.005&0.048&-0.16\\
144402&7.312$\pm$0.008&-0.052$\pm$0.005&0.063&-0.40\\
184415&7.247$\pm$0.008&0.014$\pm$0.007&0.058&0.08\\
209181&7.263$\pm$0.009&-0.018$\pm$0.008&0.072&0.03\\
209743&7.284$\pm$0.007&-0.029$\pm$0.009&0.057&-0.21\\
220439&7.209$\pm$0.006&0.026$\pm$0.007&0.052&0.12\\
227970&7.230$\pm$0.012&0.063$\pm$0.011&0.095&0.25\\
238395&7.291$\pm$0.007&0.013$\pm$0.006&0.065&0.07\\
273952&7.306$\pm$0.010&0.066$\pm$0.011&0.086&0.29\\
279818&7.299$\pm$0.008&0.048$\pm$0.010&0.072&0.11\\
422366&7.303$\pm$0.014&0.006$\pm$0.020&0.109&0.01\\
463288&7.370$\pm$0.015&-0.006$\pm$0.018&0.135&0.03\\
487027&7.205$\pm$0.004&-0.033$\pm$0.006&0.043&-0.12\\
492414&7.312$\pm$0.007&0.032$\pm$0.007&0.059&0.09\\
610997&7.239$\pm$0.011&-0.025$\pm$0.012&0.087&-0.03\\
618116&7.229$\pm$0.006&0.010$\pm$0.006&0.055&0.09\\
622744&7.581$\pm$0.006&-0.144$\pm$0.005&0.062&-0.73\\
\end{tabular}
\caption{List of ionization parameter gradients and intercepts with their 1$\sigma$ uncertainties, root mean square (RMS) scatter and Pearson correlation coefficient (PCC) values.}
\label{iontable}
\end{table} 
\section{Discussion}
\subsection{Metallicity Gradients}
Using the KK04 \rtt strong line emission diagnostic, we find a weak dependence in the slope of the normalised radial metallicity gradient with the stellar mass of the galaxy. This is inconsistent with the results found by several other recent studies on radial metallicity gradients in galaxies \citep{sanchez12,sanchez14,ho15,sanchez-menguiano16}. However, as demonstrated by \citet{kewley08}, the calculated metallicities are strongly dependent upon the calibration used. Based on this, the derived metallicity gradients may also depend upon the particular diagnostic used. \citet{belfiore17} calculated metallicity gradients for galaxies using a diagnostic derived from the \rtt line ratio and also found a dependence on metallicity gradients with stellar mass. Although \citet{belfiore17} use the same \rtt diagnostic, they use the \citet{maiolino08} calibration to determine metallicites, making a direct comparison between results difficult. We find a mean metallicity gradient value of -0.12 dex/R$_{e}$ with a standard deviation of 0.05 using the KK04 \rtt metallicity diagnostic.
\par
Since there appears to be a dependence on metallicity gradients with stellar mass, sample selection plays an important role in the determination of mean metallicity gradients. Although \citet{belfiore17} uses a different metallicity calibration to the \rtt diagnostic, we note that they seem to find a shallower, although still consistent, mean metallicity gradient ($-0.08\pm0.12$ dex/R$_{e}$) than the ones determined here. The shallower mean metallicity gradient is caused by differences in sample selection. \citet{belfiore17} sample a wider stellar mass range, including relatively more low mass galaxies. Since metallicity gradients have a stellar mass dependence, these lower mass galaxies have shallower metallicity gradients and hence decrease the mean metallicity gradient of the sample. This effect is also demonstrated by \citet{belfiore17} with a shallower volume-limited mean metallicity gradient, where low mass galaxies are relatively heavier weighted.
\par
\citet{sanchez12,sanchez14,sanchez-menguiano16} used the PP04 O3N2 diagnostic with their sample of CALIFA galaxies in order to analyse the metallicity gradients of galaxies and \citet{sanchez-menguiano16} found a mean metallicity gradient of $-0.11 \pm 0.07$ dex/R$_{e}$. We recalculate our gradients using the PP04 O3N2 diagnostic and find a mean metallicity gradient of $-0.10 \pm 0.06$ dex/R$_{e}$ after excluding the inner (R/R$_{e}<0.5$) and outer sections (R/R$_{e}>2.0$) of the galaxies in the same way as \citet{sanchez12,sanchez14,sanchez-menguiano16}. Our metallicity gradients are consistent with those presented in all three studies. Our results are also consistent with the PP04 O3N2 metallicity gradients presented in \citet{belfiore17}, which found a mean metallicity gradient of $-0.08 \pm 0.10$ dex/R$_{e}$. \citet{belfiore17} again presents with slightly shallower but still consistent mean metallicity gradient. \citet{belfiore17} also finds a mass dependence of the O3N2 metallicity gradients, meaning their wider stellar mass range may explain their slightly shallower mean metallicity gradient. 
\par
For metallicity diagnostics which display mass-dependent metallicity gradients, sample selection appears to have a strong influence on the calculated mean metallicity gradient. Therefore, care must be taken when comparing results between different studies as the stellar mass distribution of the sample may have a heavy impact on the results obtained.
\par
\citet{sanchez12,sanchez14} and \citet{sanchez-menguiano16} excluded the inner (R/R$_{e}<0.5$) and outer (R/R$_{e}$ > 2.0) galactic radii when measuring the metallicity gradients because of the observed flattening of the metallicity gradient that occurs at these radii \citep{bresolin09,bresolin12,rosales-ortega11,marino12,sanchez12,sanchez14,sanchez-menguiano16}. We find metallicity gradient flattening occurring at R/R$_{e}<0.5$ only for GAMA-106717 using either the PP04 O3N2 or KK04 diagnostic. Only two of our galaxies (GAMA-144402 and GAMA-622744) are observed beyond 2R$_{e}$, and neither show any clear flattening of the metallicity gradient.

\par
\citet{ho15} used the KD02 metallicity diagnostic to determine the metallicity gradients of a sample of CALIFA and WiFeS galaxies. Using the R$_{25}$ scale length to normalise the metallicity gradients, \citet{ho15} found no significant dependence on stellar mass. \citet{ho15} found a mean metallicity gradient of $-0.39 \pm 0.18$ dex/R$_{25}$. We determine the metallicity gradients using the KD02 diagnostic, but the uncertainties in R$_{25}$ for our sample were too large for a reliable comparison (based on values obtained from HyperLeda \citep{hyperleda}). We instead assume a crude approximation of R$_{25}$=3.6R$_{e}$ based on fits to S0 galaxies by \citet{williams09}. Using this approximation, we obtain a mean metallicity gradient of $-0.48 \pm 0.18$ dex/R$_{25}$. Although the metallicity gradients agree within the errors, it is important to keep in mind that we have only used an approximation to R$_{25}$ and have used the R-band scale length instead of the B-band which was used in \cite{ho15}.
\subsection{Scatter around Metallicity Gradients}
For the majority of metallicity gradients, the scatter increases noticeably at larger radii. Within 1R$_{e}$, the standard deviation away from the metallicity gradient is approximately 0.04 dex and and increases to 0.08 dex beyond 1Re. We find that this is driven mostly by the decrease in line flux, and hence S/N, at larger radii in the SAMI data. At an integrated H$\alpha$ S/N < 80 the scatter is 0.07 dex, whereas at a H$\alpha$ S/N > 80 the scatter decreases to about 0.03 dex. However, a decrease in S/N does not account for all of the increase in scatter. 
\par
In five of our galaxies we notice that more than half of the spaxels within 1R$_{e}$ have a S/N < 80, and have significantly less scatter than those spaxels at radii larger than 1R$_{e}$. We also notice a large bias of the scatter towards lower metallicities. We find that spaxels which deviate more than 0.1dex from the metallicity gradient have an increased \rtt line ratio. All these spaxels also lie on the upper branch of the KK04 \rtt metallicity diagnostic. The combination of these two effects leads to a lower metallicity measurement. In addition, the KK04 \rtt metallicity diagnostic becomes less sensitive to metallicity at higher values of \rtt, which only enhances this deviation. 
\par
The larger \rtt line ratio is caused by an increase in both the [OII]/H$\beta$ and [OIII]/H$\beta$ line ratios. The [OII]/H$\beta$ line ratio has a larger percentage increase than the [OIII]/H$\beta$ line ratio. This leads to an overall decrease in the [OIII]/[OII] line ratio, causing lower ionization parameter measurements for a metallicity range of $7.6<$12$+$log(O/H)$<9.2$. 
\par
One explanation for the enhanced line ratios at large radii is diffuse ionized gas (DIG) contamination. Using data from the MaNGA survey, \citet{zhang17} demonstrated the effects of DIG on emission line ratios and metallicity diagnostics. They found that the [OII]/H$\beta$ line ratio is enhanced in DIG dominated regions, while the DIG effects on the [OIII]/H$\beta$ line ratio depends on the specific situation of the galaxy. In both cases, they also found a decrease in the [OIII]/[OII] line ratio.
\subsection{Mass-Metallicity Relation}
Figure \ref{metint} shows the metallicity intercepts as a function of stellar mass with the mass-metallicity fit from \citet{kewley08} shown as the dotted red line. A small positive offset of 0.13 was required to optimally fit the mass-metallicity relation to the intercept data, this is shown as the solid red line. This is to account for the fact that we are using the interpolated central metallicity which simulates an infinitesimally small central aperture. The interpolated central metallicity would be systematically higher than the global metallicity or larger aperture metallicity measurements because we are not averaging the regions of high and low metallicity. \citet{tremonti04} was able to simulate the effects of changing aperture metallicity measurements by showing that nearer galaxies had larger aperture metallicities than those further away of similar size. The nearer galaxies had a larger apparent size, meaning that they were restricted to sampling a smaller fraction of the galaxy.
\subsection{Ionization Parameter Gradients}
The ionization parameter maps produced by \citet{kaplan16} show significant radial ionization parameter gradients as well as a correlation with SFR. Correlation between ionization parameter and SFR was also observed by \citet{dopita14} using a sample of WiFeS galaxies.
\par
\citet{yuan13,mast14} have shown that decreased spatial resolution leads to the flattening of observed metallicity gradients. \citet{kaplan16} has a median resolution of 387pc, while the full SAMI survey has resolutions on the order of kpc. Our galaxy sample has a median spatial resolution of 1.3kpc/PSF caused by the seeing limited observations with an average seeing of $2.16\arcsec$ and DAR smoothing of $0.8\arcsec$. Our galaxies are significantly less massive and have a higher redshift, meaning that fine details are difficult to resolve compared to \citet{kaplan16}. It is possible that the lack of ionization parameter gradients is due to the spatial smoothing caused by our inability to resolve the finer details due to limitations in seeing. More work using higher resolution data is needed in order to confirm if ionization parameter gradients are affected in the same way as metallicity gradients.
\par
The SAMI spectrograph does not have the spatial resolution required to resolve \htwo regions at the redshift of the main galaxy survey. To obtain higher spatial resolution spectra of \htwo regions, a sister survey of nearby \htwo regions is being conducted in order to recalibrate the strong line emission diagnostics (SAMI Zoom, Sweet et al. In Prep). 
\par
The galaxies used in \citet{kaplan16} are also more massive ($10.2<\log(M_{*}/M_{\odot})<11.6$) than the mass range of the galaxies ($9.0<\log(M_{*}/M_{\odot})<10.5$) used in this study. Although we find no variation in ionization parameter gradient or intercept with mass, the difference in galaxy masses could be a factor in the absence of ionization parameter gradients.
\subsection{Ionization Parameter and Galaxy Properties}
\citet{dopita14} quantified the relationship between ionization parameter and SFR[M$_{\odot}$/year/kpc$^{2}$] as $q$[cm/s]$ \propto \textrm{SFR[M$_{\odot}$/year/kpc$^2$]}^{0.34\pm0.08}$ when $\log(q$[cm/s]$) \gtrsim 7.2-7.4$. From Figure 13 of \citet{dopita14}, we observe that below $\log(\textrm{SFR[M$_{\odot}$/year/kpc$^2$]}) < -0.5$, the correlation disappears and no trends are observed. Figure \ref{SFRiongradgrid} shows that all of our spaxels lie below $\log(\textrm{SFR[M$_{\odot}$/year/kpc$^2$]}) < -0.5$ with the large majority below $\log(\textrm{SFR[M$_{\odot}$/year/kpc$^2$]}) < -1.0$. We believe that this is the main reason that we do not observe the same trends as \citet{dopita14}. The sSFR is even less correlated with ionization parameter, with PCC values consistently lower than those of SFR. 
\par
GAMA-622744 is the only galaxy that displays a significant correlation between metallicity and ionization parameter. However we believe this is not necessarily caused by an intrinsic relationship between metallicity and ionization parameter, but rather because GAMA-622744 is the only galaxy which possess a significant ionization parameter gradient. The positive correlation contradicts the theoretical relation presented in \citet{dopita06} ($q$[cm/s]$ \propto Z$[O/H]$^{-0.8}$). Many of the galaxies in \citet{dopita14} show a positive correlation between ionization parameter and metallicity while our work lacks any significant trends.
\subsection{Ionization Parameter effects on Metallicity Diagnostics}
In Figure \ref{iongradgrid}, we see that the typical ionization parameter range for our galaxy sample is $7.0 < \log(q$[cm/s]$) < 7.8$. An ionization parameter range this wide is enough to significantly affect the metallicity estimates for several metallicity diagnostics \citep{kewley02}. As there are no discernible patterns in the distribution of ionization parameter, it makes it difficult to predict how the exclusion of ionization parameter will affect the metallicity distribution. We advise caution when interpreting results which have used metallicity diagnostics where ionization parameter has not been taken into account.

\section{Summary}

We have presented metallicity and ionization parameter maps of 25 high-S/N face-on star-forming galaxies in DR1 of the SAMI galaxy survey. To account for their interdependence, metallicity and ionization parameter were determined simultaneously for individual spaxels using an iterative method involving the strong emission line diagnostics outlined in \citet{kobulnicky04}.
\par
We measure metallicity gradients as a function of galactocentric radius using robust line fitting routines. We find that the majority of galaxies exhibit a negative metallicity gradient with an average metallicity gradient of -0.12$\pm$0.05 dex/R$_{e}$ using the KK04 \rtt diagnostic. Metallicity gradients show a weak negative correlation with the stellar mass of galaxies.
\par
Using the PP04 O3N2 metallicity diagnostic we find an average metallicity gradient of -0.10$\pm$0.06 dex/Re, which agrees with the gradients determined by \citet{sanchez12,sanchez14,sanchez-menguiano16} and \citet{belfiore17}. Due to the unreliable R$_{25}$ measurements of the galaxies in our sample, we are unable to directly compare our metallicity gradient value to \citet{ho15}. However, assuming $R_{25}=3.6$R$_{e}$ based on \citet{williams09}, we find an average N2O2 metallicity gradient of $-0.48 \pm 0.18$, consistent with that of \citet{ho15}.
\par
Using the central metallicities of each galaxy based on the linear fits, we find that our galaxies are in agreement with the mass-metallicity relation polynomial presented in \citet{kewley08} after applying a positive offset of 0.13 dex. The offset is likely a result of using interpolated central metallicities rather than the aperture average value as determined for SDSS.
\par
We show that the ionization parameter maps lack significant or coherent structure unlike the metallicity maps. We do not see significant ionization parameter gradients like those presented in \citet{kaplan16}, however this could be due to sample selection differences or spatial resolution limitations. We do find a decrease in ionization parameter in the inter-arm regions of galaxies with resolvable spiral arms indicating a possible correlation between ionization parameter and SFR. However for our galaxy sample, we find no significant correlations between ionization parameter and SFR or sSFR.
\par
Until a better understanding is achieved on the distribution of ionization parameter, metallicity diagnostics must be used with care. We suggest that in order to obtain reliable metallicity maps, to either use a metallicity diagnostic which explicitly provides solutions for a range of ionization parameter like the one used in this study (eg. KK04 R23), or use a metallicity diagnostic which is relatively invariant to changes in ionization parameter (eg. KD02 N2O2 or D16 N2S2). 
\section*{Acknowledgements}

The SAMI Galaxy Survey is based on observations made at the Anglo-Australian Telescope. The Sydney-AAO Multi-object Integral field spectrograph (SAMI) was developed jointly by the University of Sydney and the Australian Astronomical Observatory. The SAMI input catalogue is based on data taken from the Sloan Digital Sky Survey, the GAMA Survey and the VST ATLAS Survey. The SAMI Galaxy Survey is funded by the Australian Research Council Centre of Excellence for All-sky Astrophysics (CAASTRO), through project number CE110001020, and other participating institutions. The SAMI Galaxy Survey website is http://sami-survey.org/.
\\
Parts of this research were conducted by the Australian Research Council Centre of Excellence for All Sky Astrophysics in 3 Dimensions (ASTRO 3D), through project number CE170100013.
\\
BG gratefully acknowledges the support of the Australian Research Council as the recipient of a Future Fellowship (FT140101202).
\\
Support for AMM is provided by NASA through Hubble Fellowship grant \#HST-HF2-51377 awarded by the Space Telescope Science Institute, which is operated by the Association of Universities for Research in Astronomy, Inc., for NASA, under contract NAS5-26555.
\\
JvdS is funded under Bland-Hawthorn's ARC Laureate Fellowship (FL140100278).
\\
SFS thanks CONACYT programs CB-285080 and DGAPA-PAPIIT IA101217 grants for their support to this project.
\\
SB acknowledges the funding support from the Australian Research Council through a Future Fellowship (FT140101166).
\\
MSO acknowledges the funding support from the Australian Research Council through a Future Fellowship (FT140100255).
\\
NS acknowledges support of a University of Sydney Postdoctoral Research Fellowship.

\bibliographystyle{mnras}
\bibliography{References}

\appendix

\section{Metallicity Maps}
\begin{landscape}
	\begin{figure}
	\centering
	\includegraphics[width=1.3\textheight]{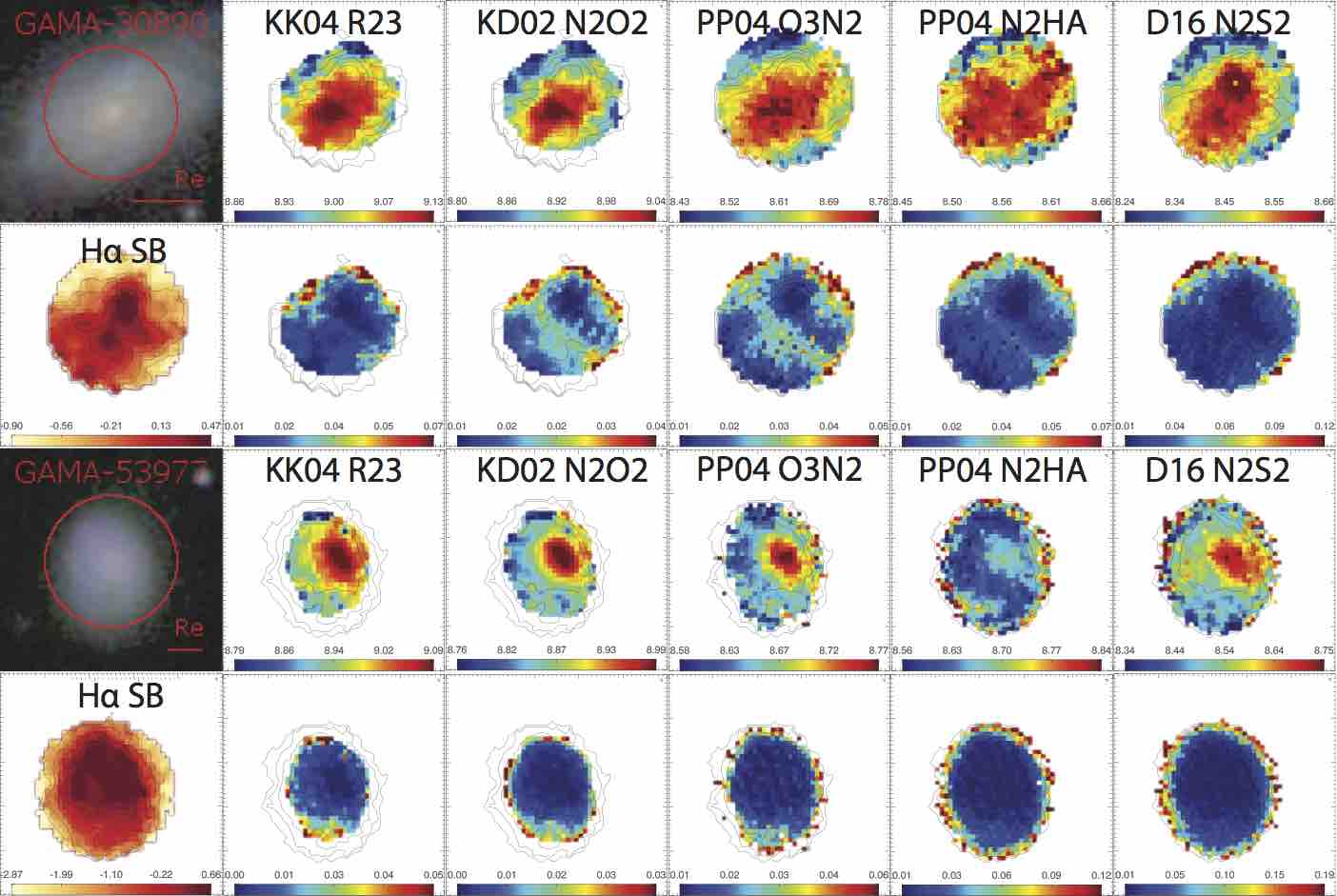}\label{fig:a}
	\caption{
Same as Figure \ref{metmapgrid} for GAMA-30890 and GAMA-53977.
	}
	\end{figure}
\end{landscape}
\begin{landscape}
	\begin{figure}
	\centering
	\includegraphics[width=1.3\textheight]{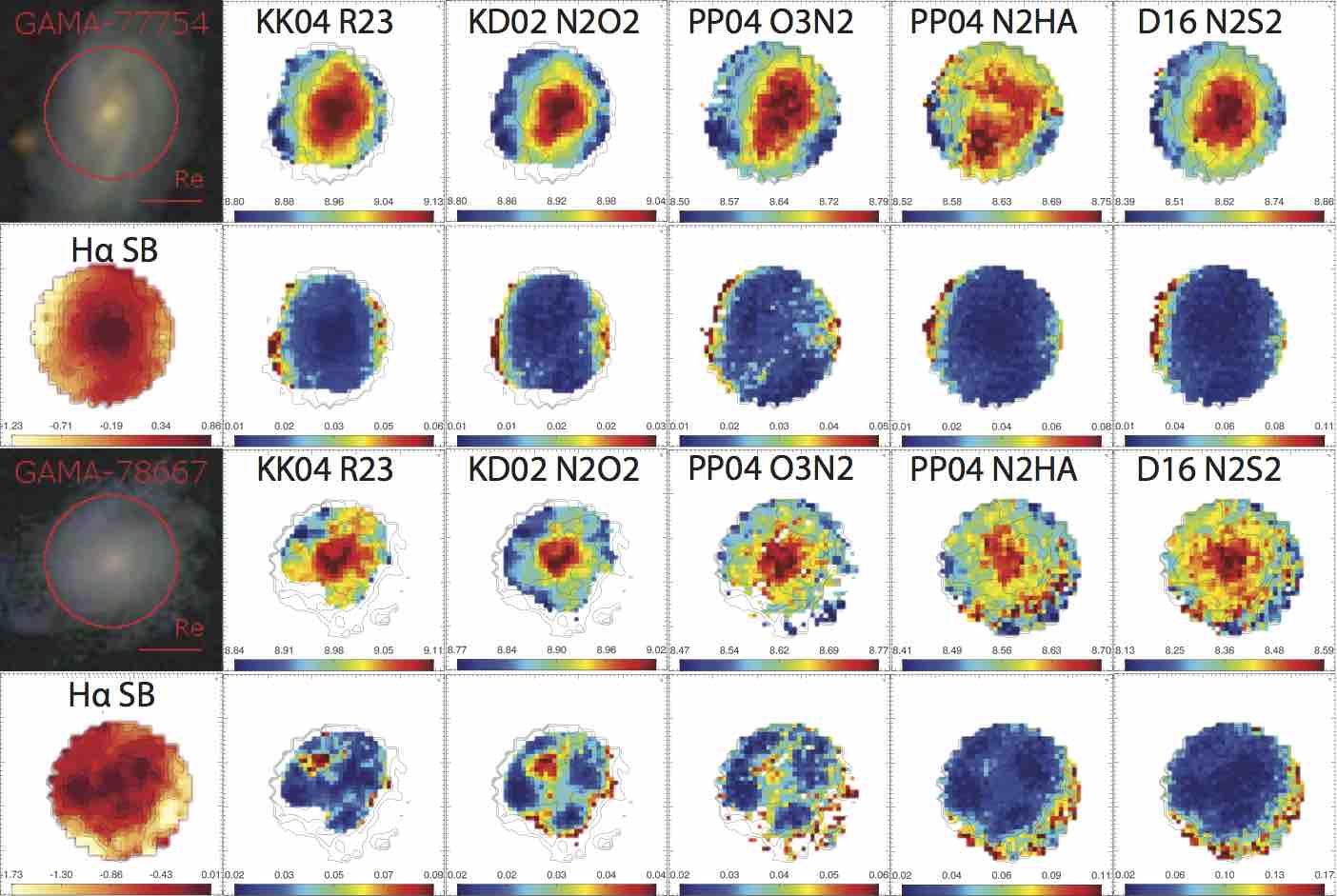}\label{fig:a}
	\caption{
Same as Figure \ref{metmapgrid} for GAMA-77754 and GAMA-78667.
	}
	\end{figure}
\end{landscape}
\begin{landscape}
	\begin{figure}
	\centering
	\includegraphics[width=1.3\textheight]{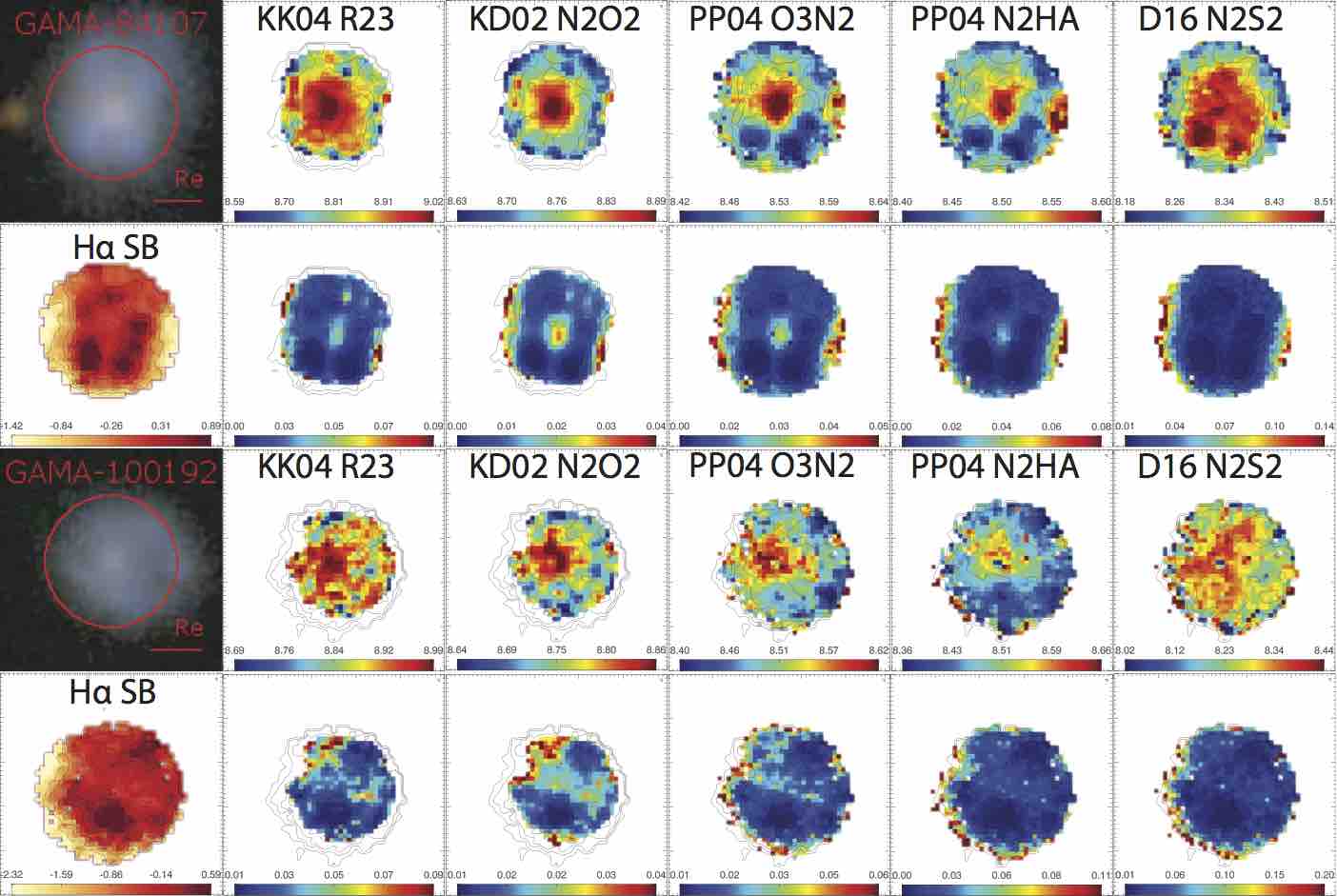}\label{fig:a}
	\caption{
Same as Figure \ref{metmapgrid} for GAMA-84107 and GAMA-100192.
	}
	\end{figure}
\end{landscape}
\begin{landscape}
	\begin{figure}
	\centering
	\includegraphics[width=1.3\textheight]{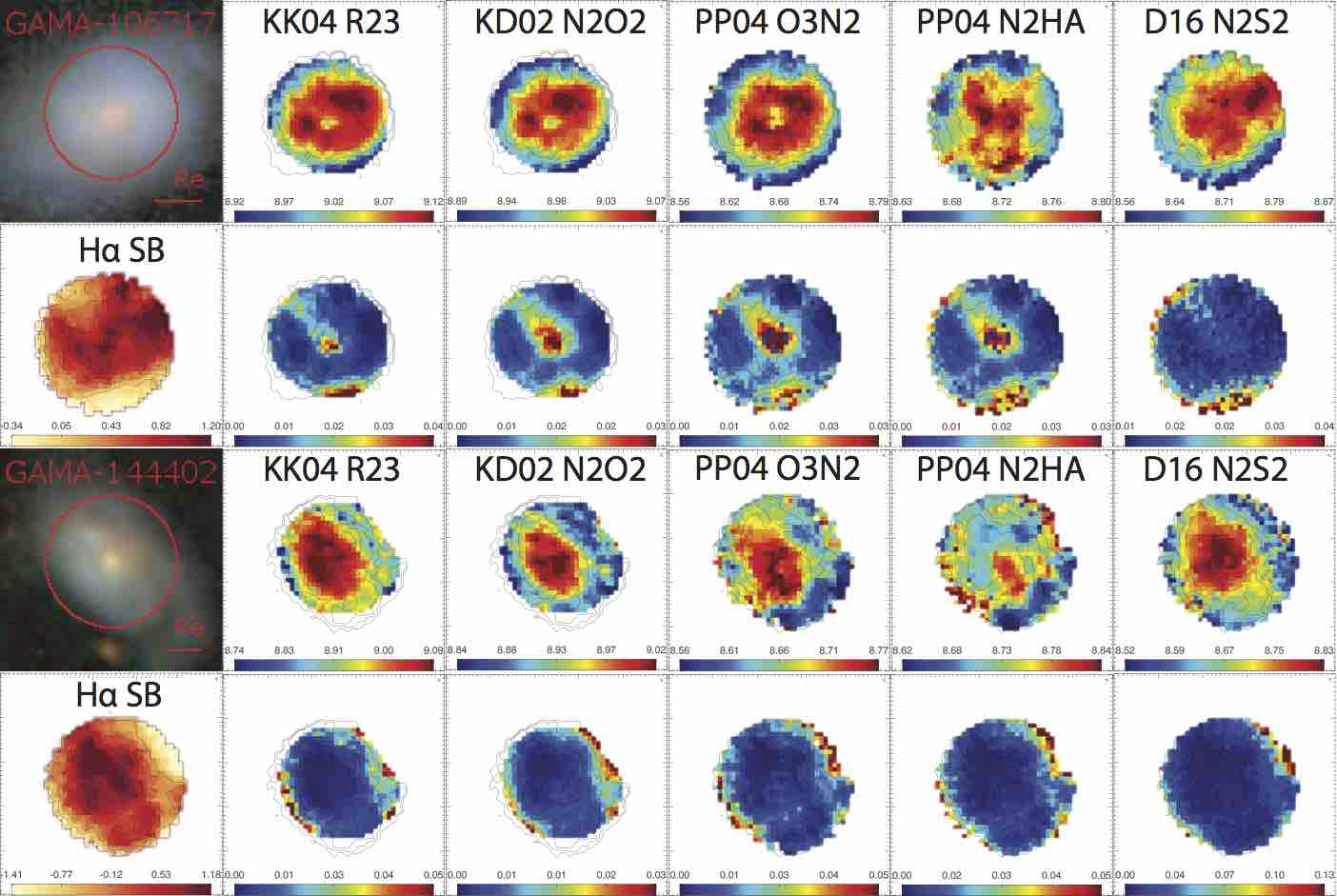}\label{fig:a}
	\caption{
Same as Figure \ref{metmapgrid} for GAMA-106717 and GAMA-144402.
	}
	\end{figure}
\end{landscape}
\begin{landscape}
	\begin{figure}
	\centering
	\includegraphics[width=1.3\textheight]{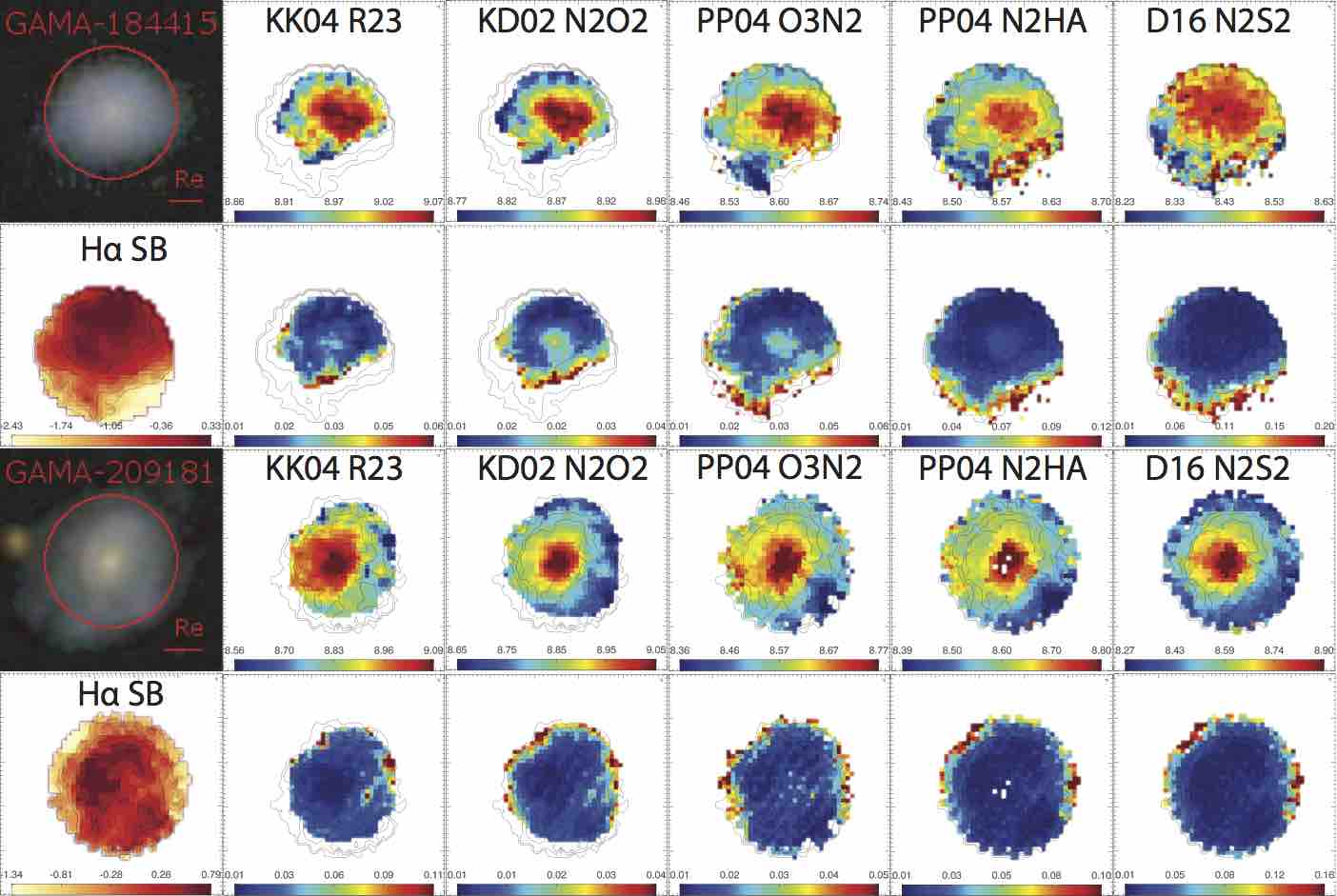}\label{fig:a}
	\caption{
Same as Figure \ref{metmapgrid} for GAMA-184415 and GAMA-209181.
	}
	\end{figure}
\end{landscape}
\begin{landscape}
	\begin{figure}
	\centering
	\includegraphics[width=1.3\textheight]{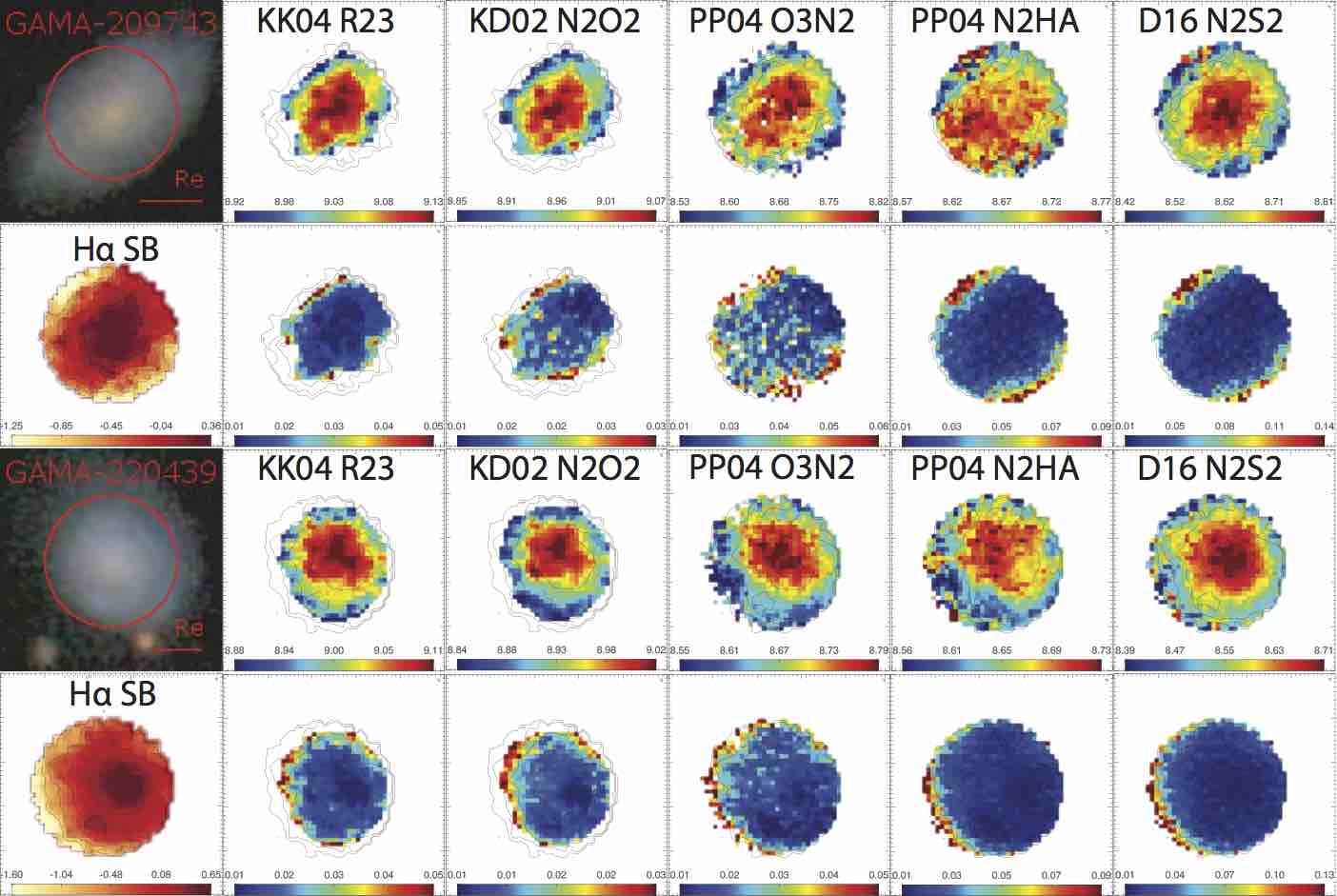}\label{fig:a}
	\caption{
Same as Figure \ref{metmapgrid} for GAMA-209743 and GAMA-220439.
	}
	\end{figure}
\end{landscape}
\begin{landscape}
	\begin{figure}
	\centering
	\includegraphics[width=1.3\textheight]{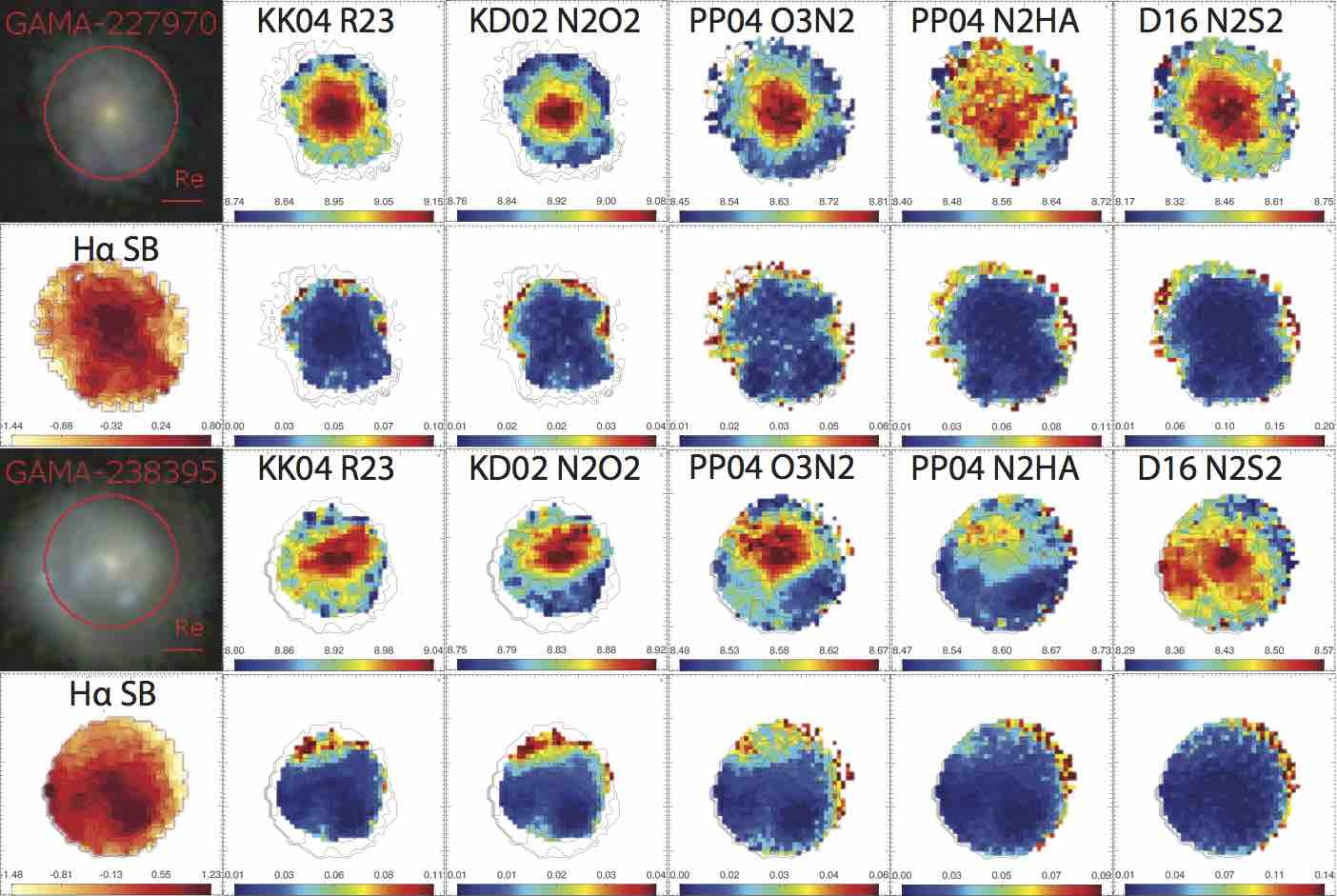}\label{fig:a}
	\caption{
Same as Figure \ref{metmapgrid} for GAMA-227970 and GAMA-238395.
	}
	\end{figure}
\end{landscape}
\begin{landscape}
	\begin{figure}
	\centering
	\includegraphics[width=1.3\textheight]{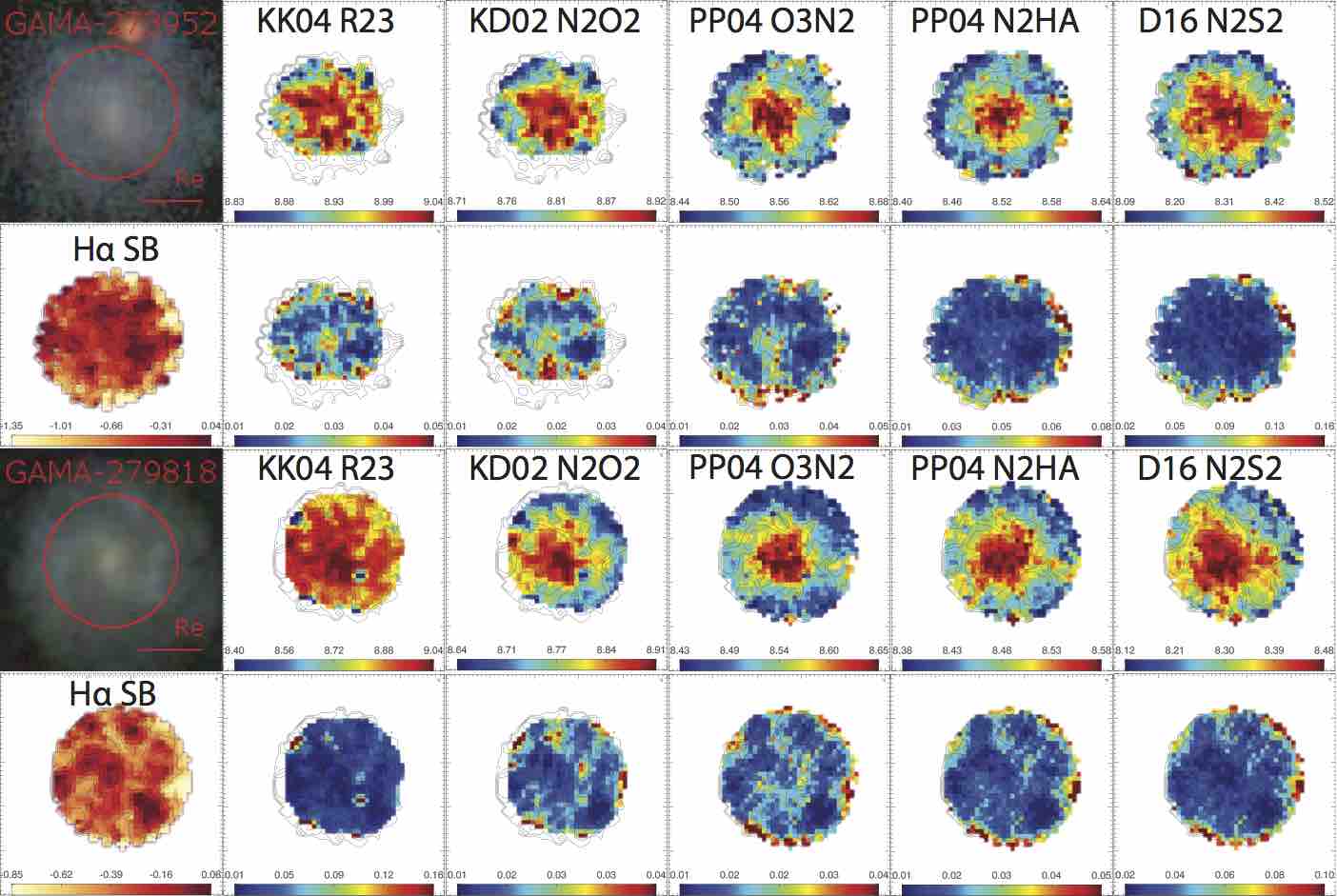}\label{fig:a}
	\caption{
Same as Figure \ref{metmapgrid} for GAMA-273952 and GAMA-279818.
	}
	\end{figure}
\end{landscape}
\begin{landscape}
	\begin{figure}
	\centering
	\includegraphics[width=1.3\textheight]{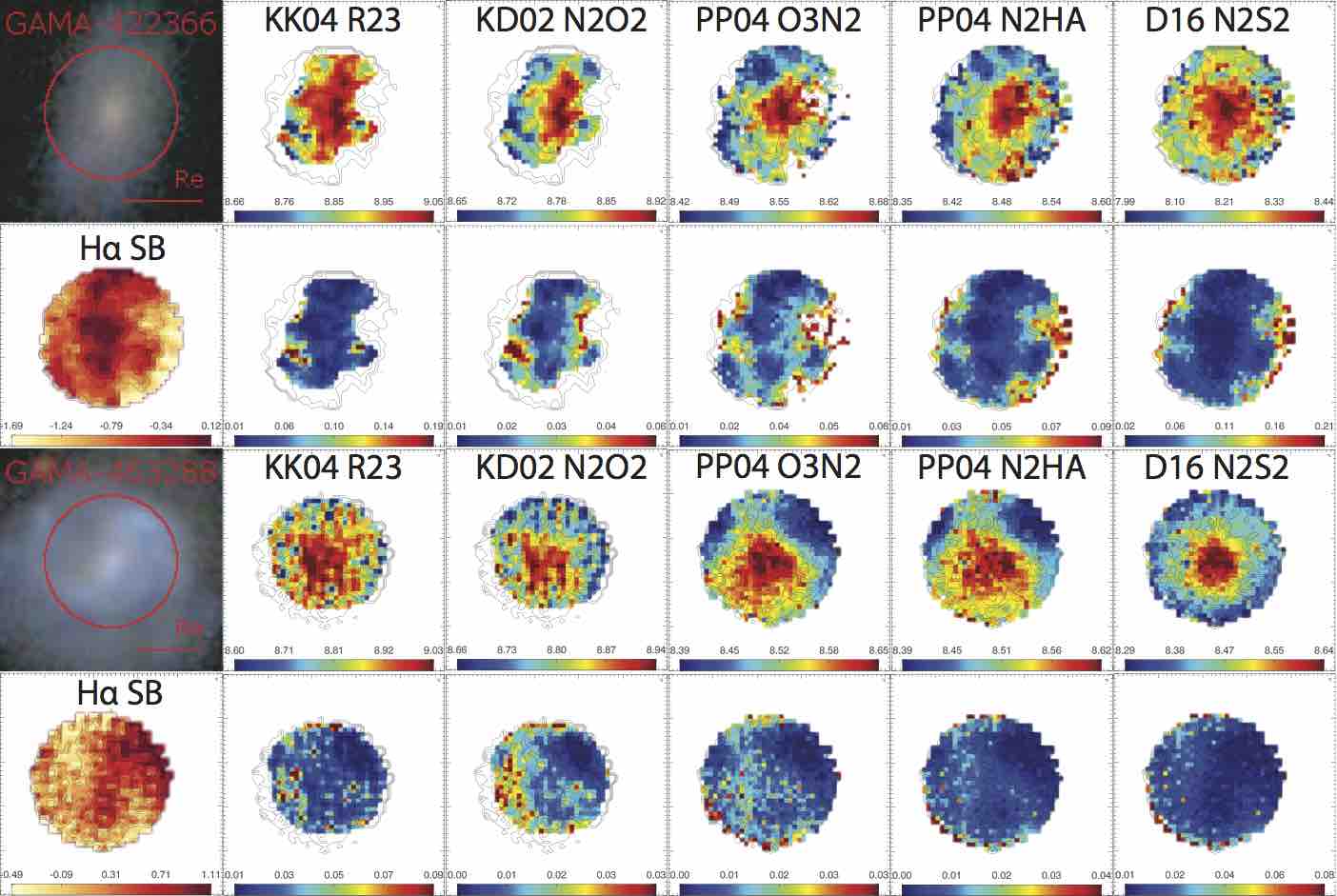}\label{fig:a}
	\caption{
Same as Figure \ref{metmapgrid} for GAMA-422366 and GAMA-463288.
	}
	\end{figure}
\end{landscape}
\begin{landscape}
	\begin{figure}
	\centering
	\includegraphics[width=1.3\textheight]{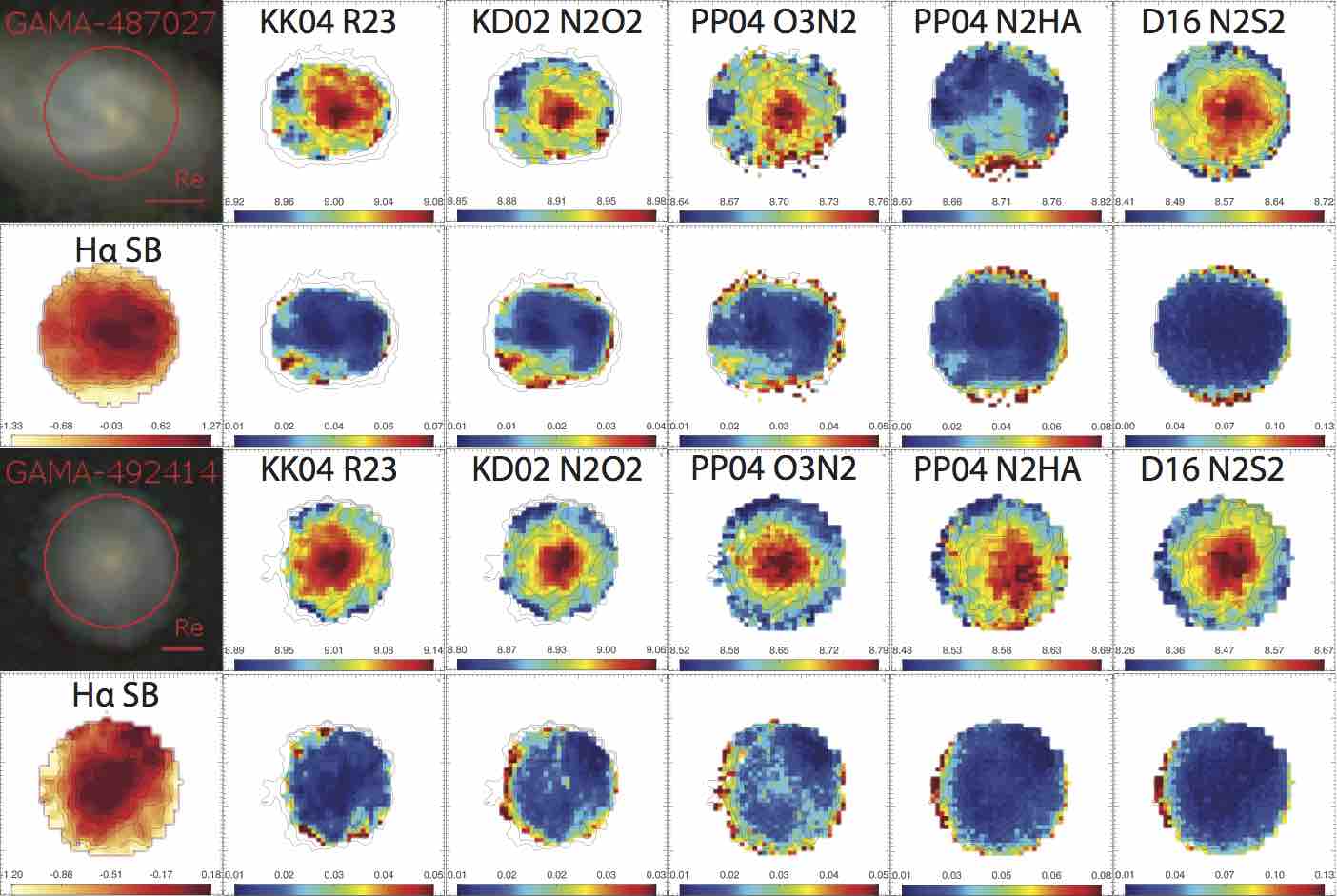}\label{fig:a}
	\caption{
Same as Figure \ref{metmapgrid} for GAMA-487027 and GAMA-492414.
	}
	\end{figure}
\end{landscape}
\begin{landscape}
	\begin{figure}
	\centering
	\includegraphics[width=1.3\textheight]{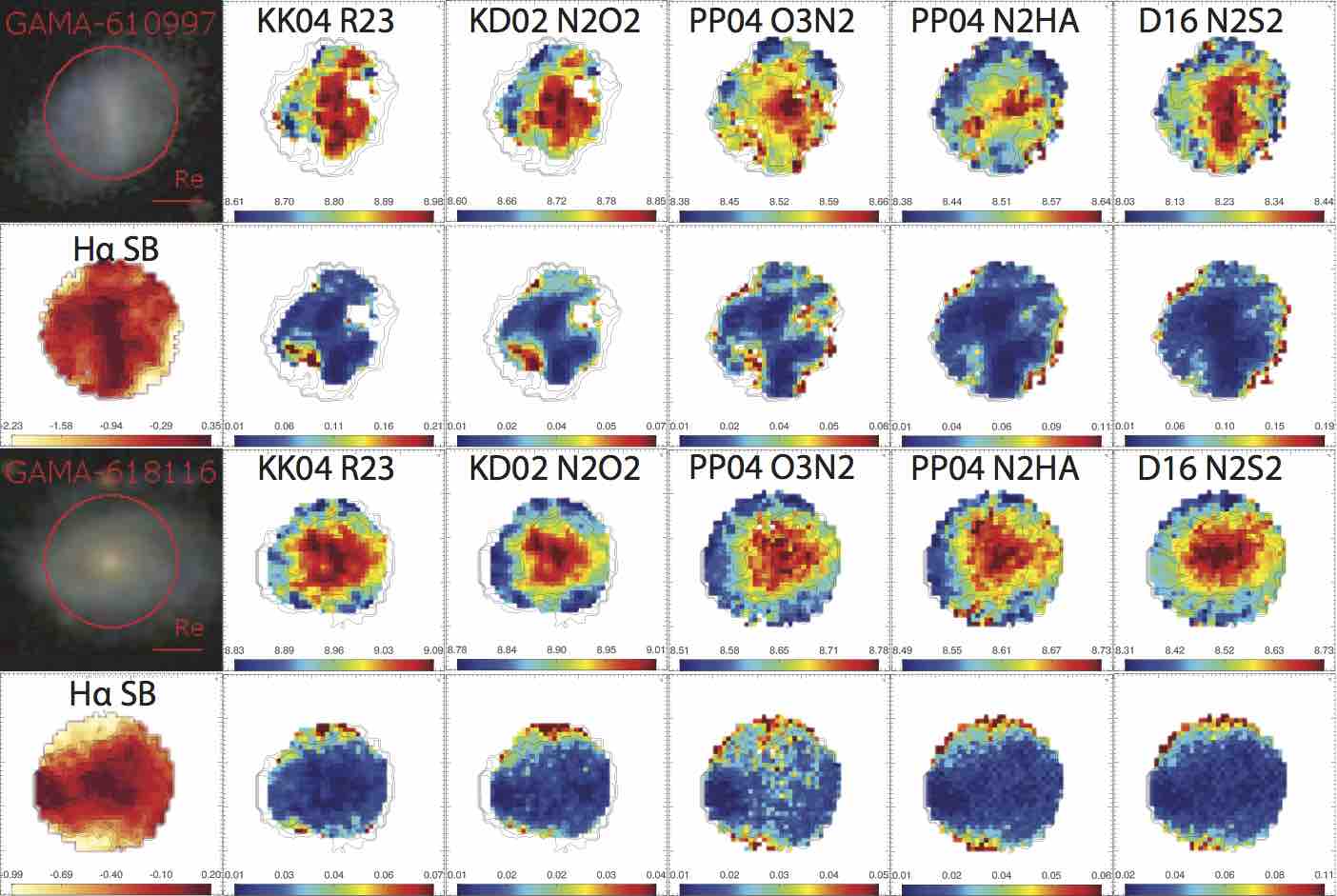}\label{fig:a}
	\caption{
Same as Figure \ref{metmapgrid} for GAMA-610997 and GAMA-618116.
	}
	\end{figure}
\end{landscape}
\section{Ionization Parameter Maps}
\begin{landscape}
	\begin{figure}
	\centering
	\includegraphics[width=1.3\textheight]{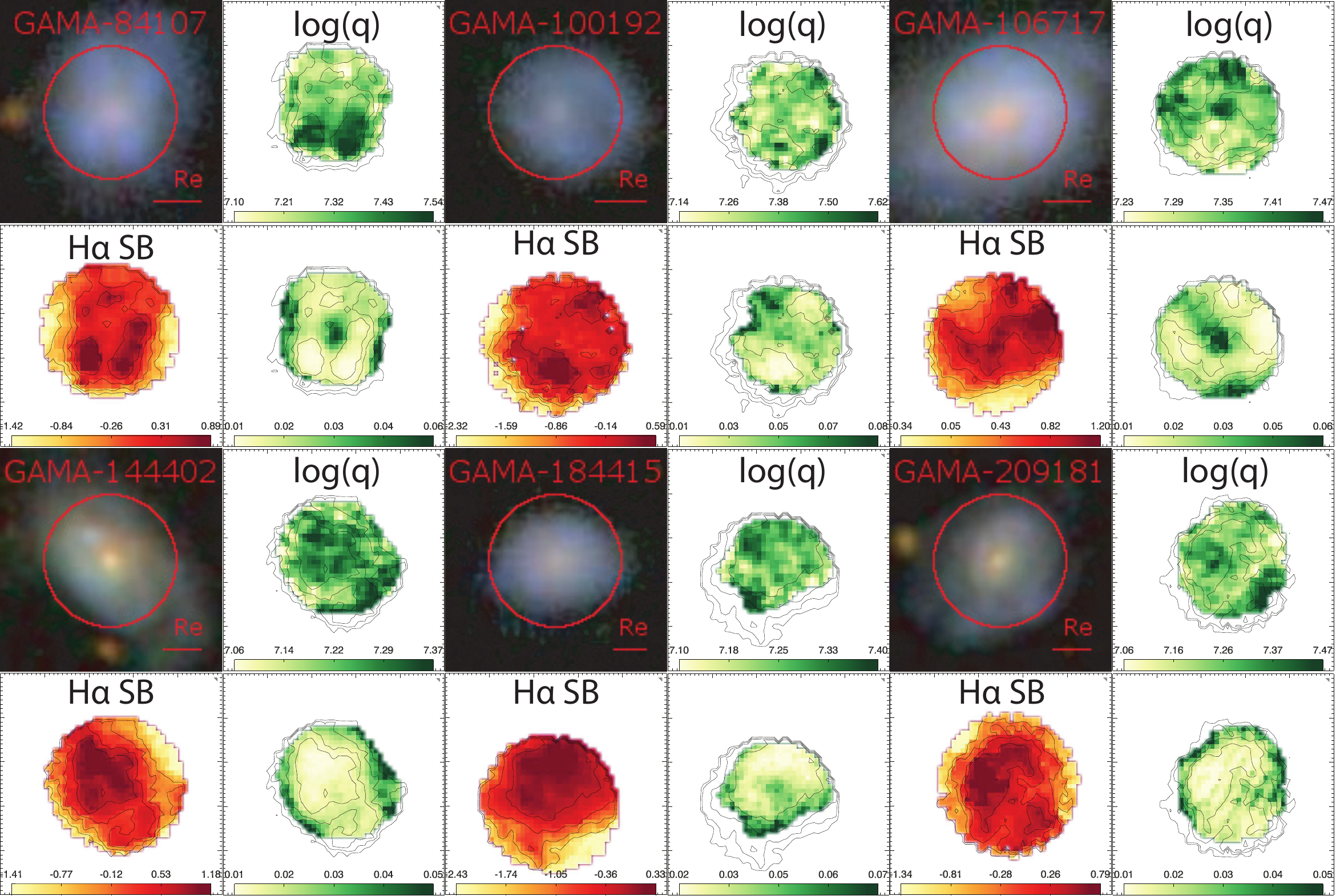}\label{fig:a}
	\caption{
Same as Figure \ref{ionmapgrid} for GAMA-84107, GAMA-100192, GAMA-106717, GAMA-144402, GAMA-184415 and GAMA-209181.
	}
	\end{figure}
\end{landscape}
\begin{landscape}
	\begin{figure}
	\centering
	\includegraphics[width=1.3\textheight]{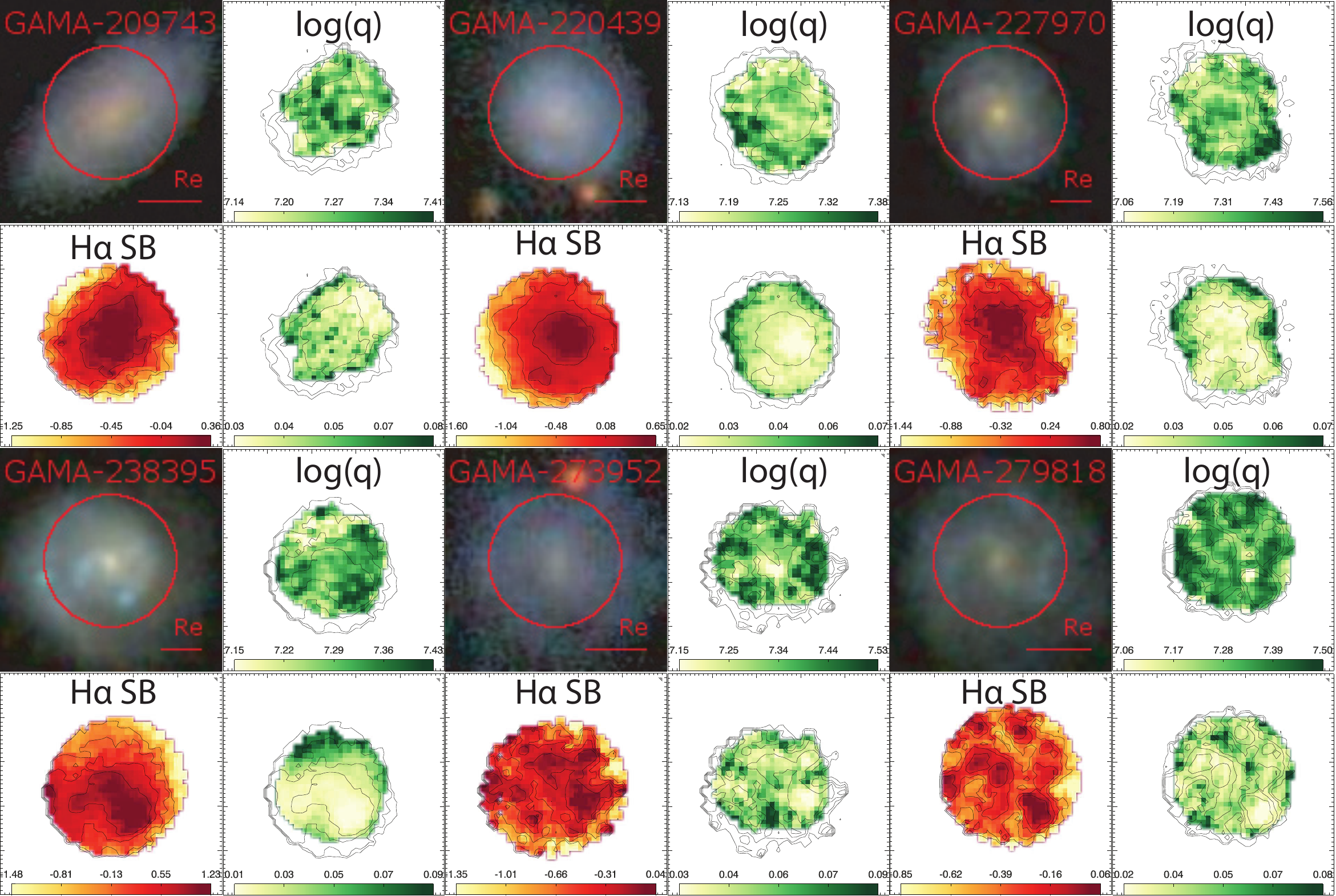}\label{fig:a}
	\caption{
Same as Figure \ref{ionmapgrid} for GAMA-209743, GAMA-220439, GAMA-227970, GAMA-238395, GAMA-273952 and GAMA-279818.
	}
	\end{figure}
\end{landscape}
\begin{landscape}
	\begin{figure}
	\centering
	\includegraphics[width=1.3\textheight]{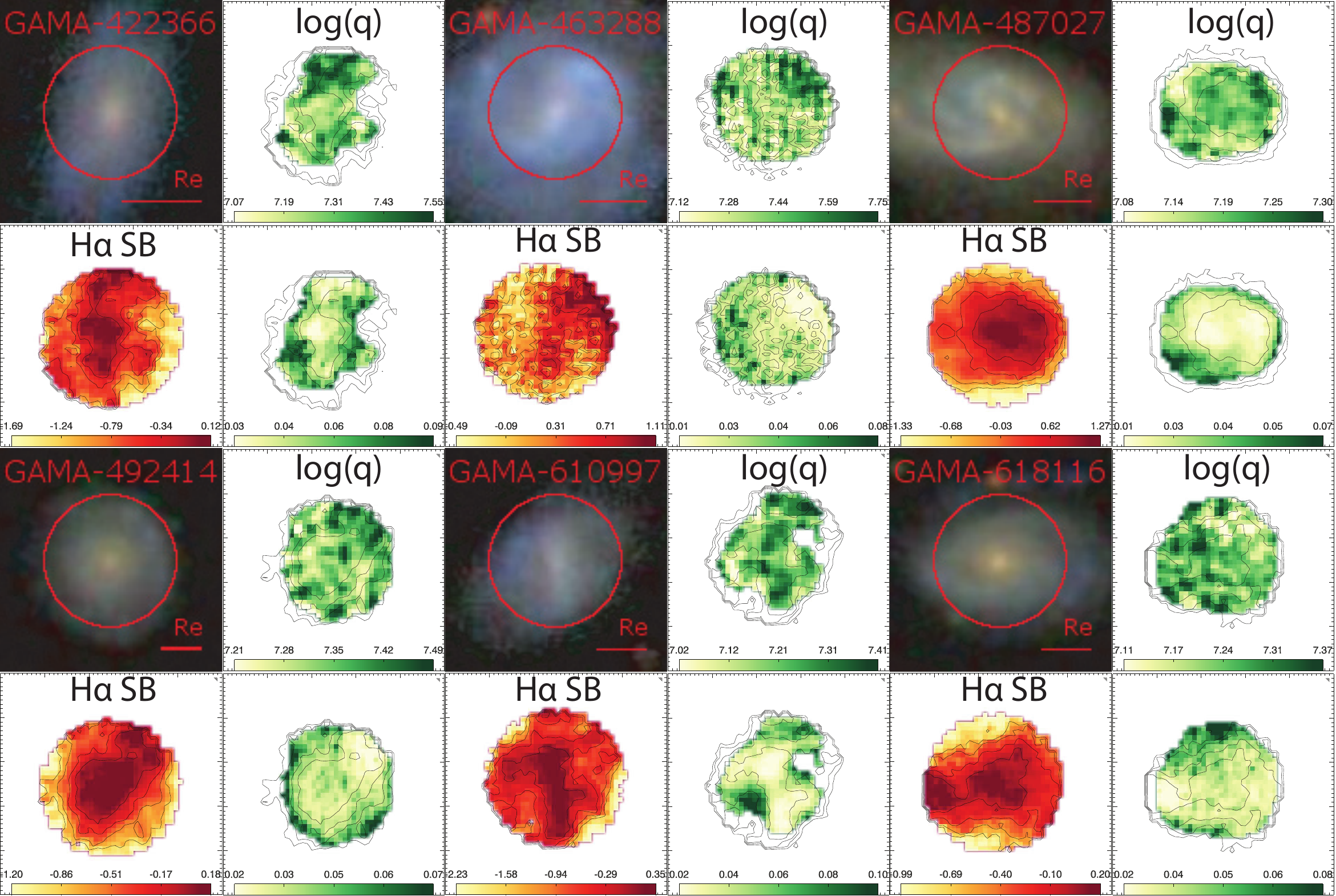}\label{fig:a}
	\caption{
Same as Figure \ref{ionmapgrid} for GAMA-422366, GAMA-463288, GAMA-487027, GAMA-492414, GAMA-610997 and GAMA-618116.
	}
	\end{figure}
\end{landscape}
\section{Tables}
\clearpage
\renewcommand{\tabcolsep}{0.15cm}
\begin{table}
\centering
\begin{tabular}{l*{5}{c}}
GAMA ID & Intercept & Gradient & RMS & PCC\\
&	12+log(O/H)&	dex/$\log$(SFR)& \\
\hline
008353&8.837$\pm$0.017&0.037$\pm$0.010&0.060&0.22\\
022633&9.273$\pm$0.016&0.148$\pm$0.010&0.075&0.48\\
030890&9.334$\pm$0.013&0.146$\pm$0.007&0.046&0.71\\
053977&9.139$\pm$0.013&0.091$\pm$0.008&0.049&0.40\\
077754&9.310$\pm$0.008&0.182$\pm$0.005&0.045&0.81\\
078667&9.317$\pm$0.025&0.132$\pm$0.012&0.049&0.38\\
084107&9.010$\pm$0.015&0.074$\pm$0.008&0.074&0.34\\
100192&8.902$\pm$0.016&-0.003$\pm$0.008&0.063&0.02\\
106717&9.214$\pm$0.006&0.129$\pm$0.005&0.032&0.63\\
144402&9.149$\pm$0.007&0.113$\pm$0.004&0.046&0.71\\
184415&9.089$\pm$0.019&0.047$\pm$0.010&0.050&0.34\\
209181&9.167$\pm$0.021&0.167$\pm$0.013&0.101&0.49\\
209743&9.493$\pm$0.014&0.235$\pm$0.008&0.034&0.75\\
220439&9.296$\pm$0.011&0.147$\pm$0.006&0.037&0.65\\
227970&9.301$\pm$0.014&0.181$\pm$0.009&0.077&0.63\\
238395&8.992$\pm$0.008&0.029$\pm$0.006&0.052&0.36\\
273952&8.951$\pm$0.023&-0.010$\pm$0.010&0.047&-0.06\\
279818&8.981$\pm$0.041&0.027$\pm$0.018&0.080&0.02\\
422366&9.155$\pm$0.024&0.082$\pm$0.011&0.073&0.31\\
463288&8.890$\pm$0.017&0.002$\pm$0.010&0.085&-0.02\\
487027&9.114$\pm$0.002&0.066$\pm$0.002&0.023&0.74\\
492414&9.362$\pm$0.011&0.146$\pm$0.006&0.040&0.58\\
610997&9.051$\pm$0.021&0.076$\pm$0.010&0.077&0.35\\
618116&9.461$\pm$0.019&0.221$\pm$0.009&0.049&0.46\\
622744&8.879$\pm$0.006&0.024$\pm$0.004&0.042&0.28\\
\end{tabular}
\caption{Linear fit parameters for Figure \ref{SFRmetgradgrid}}
\label{R23SFR}
\end{table}

\begin{table}
\centering
\begin{tabular}{l*{5}{c}}
GAMA ID & Intercept & Gradient & RMS & PCC\\
&	log(q)&	dex/$\log$(SFR)& \\
\hline
008353&7.570$\pm$0.019&0.213$\pm$0.011&0.065&0.58\\
022633&7.457$\pm$0.016&0.053$\pm$0.010&0.076&0.15\\
030890&7.368$\pm$0.018&0.036$\pm$0.010&0.059&-0.02\\
053977&7.289$\pm$0.011&0.052$\pm$0.007&0.040&0.28\\
077754&7.242$\pm$0.010&-0.009$\pm$0.006&0.051&-0.08\\
078667&7.103$\pm$0.034&-0.075$\pm$0.015&0.062&-0.21\\
084107&7.602$\pm$0.018&0.166$\pm$0.010&0.084&0.53\\
100192&7.444$\pm$0.028&0.059$\pm$0.014&0.083&0.20\\
106717&7.364$\pm$0.008&0.025$\pm$0.007&0.047&0.21\\
144402&7.323$\pm$0.009&0.051$\pm$0.006&0.061&0.33\\
184415&7.396$\pm$0.023&0.071$\pm$0.012&0.054&0.02\\
209181&7.367$\pm$0.016&0.074$\pm$0.009&0.069&0.32\\
209743&7.341$\pm$0.027&0.044$\pm$0.015&0.057&0.14\\
220439&7.174$\pm$0.015&-0.032$\pm$0.008&0.053&-0.13\\
227970&7.221$\pm$0.021&-0.045$\pm$0.013&0.096&-0.12\\
238395&7.388$\pm$0.009&0.063$\pm$0.007&0.060&0.38\\
273952&7.363$\pm$0.047&0.001$\pm$0.021&0.088&-0.03\\
279818&7.437$\pm$0.041&0.046$\pm$0.019&0.074&-0.01\\
422366&7.376$\pm$0.049&0.029$\pm$0.022&0.107&0.10\\
463288&7.573$\pm$0.029&0.121$\pm$0.017&0.134&0.28\\
487027&7.238$\pm$0.006&0.046$\pm$0.005&0.043&0.23\\
492414&7.368$\pm$0.020&0.012$\pm$0.010&0.062&0.05\\
610997&7.358$\pm$0.027&0.069$\pm$0.013&0.084&0.12\\
618116&7.182$\pm$0.020&-0.028$\pm$0.010&0.055&-0.16\\
622744&7.677$\pm$0.011&0.144$\pm$0.006&0.060&0.65\\
\end{tabular}
\caption{Linear fit parameters for Figure \ref{SFRiongradgrid}}
\label{R23QSFR}
\end{table}

\begin{table}
\centering
\begin{tabular}{l*{5}{c}}
GAMA ID & Intercept & Gradient & RMS & PCC\\
&	log(q)&	dex/Z& \\
\hline
008353&2.843$\pm$0.644&0.502$\pm$0.074&0.083&0.28\\
022633&7.517$\pm$0.291&-0.016$\pm$0.033&0.079&0.01\\
030890&2.132$\pm$0.333&0.578$\pm$0.037&0.051&0.31\\
053977&2.612$\pm$0.314&0.517$\pm$0.035&0.036&0.67\\
077754&7.250$\pm$0.298&0.001$\pm$0.033&0.052&0.02\\
078667&2.256$\pm$0.436&0.563$\pm$0.049&0.054&0.33\\
084107&0.777$\pm$0.438&0.743$\pm$0.050&0.066&0.41\\
100192&1.045$\pm$0.630&0.718$\pm$0.072&0.079&0.38\\
106717&7.953$\pm$0.381&-0.069$\pm$0.042&0.049&0.11\\
144402&2.011$\pm$0.495&0.586$\pm$0.055&0.065&0.42\\
184415&6.944$\pm$0.508&0.036$\pm$0.057&0.058&0.13\\
209181&4.796$\pm$0.308&0.277$\pm$0.035&0.070&0.15\\
209743&1.378$\pm$0.455&0.655$\pm$0.051&0.052&0.51\\
220439&6.491$\pm$0.389&0.082$\pm$0.044&0.051&0.15\\
227970&6.496$\pm$0.472&0.090$\pm$0.053&0.097&0.12\\
238395&8.063$\pm$0.503&-0.086$\pm$0.057&0.065&0.02\\
273952&3.725$\pm$0.630&0.412$\pm$0.071&0.085&0.16\\
279818&4.015$\pm$0.424&0.378$\pm$0.048&0.078&0.44\\
422366&2.275$\pm$0.684&0.571$\pm$0.078&0.102&0.33\\
463288&2.297$\pm$0.705&0.576$\pm$0.080&0.131&0.30\\
487027&-1.538$\pm$0.421&0.977$\pm$0.047&0.034&0.56\\
492414&6.602$\pm$0.382&0.083$\pm$0.043&0.062&0.17\\
610997&-0.171$\pm$0.462&0.844$\pm$0.053&0.071&0.40\\
618116&5.933$\pm$0.305&0.147$\pm$0.034&0.055&0.19\\
622744&-9.987$\pm$0.721&2.011$\pm$0.083&0.068&0.73\\
\end{tabular}
\caption{Linear fit parameters for Figure \ref{metiongrid}}
\label{R23QN2O2}
\end{table}

\bsp	\label{lastpage}
\end{document}